\documentclass[tightenlines,12pt,nofootinbib,longbibliography,superscriptaddress]{revtex4-1}
\newcommand{\TheTerm}{critical drag}
\newcommand{\eqnref}[1]{Eq.~(\ref{#1})}

\usepackage[mathscr]{euscript}

\newcommand{\be}{\begin{equation}}
\newcommand{\ee}{\end{equation}}
\newcommand{\bea}{\begin{eqnarray}}
\newcommand{\eea}{\end{eqnarray}}
\newcommand{\beq}{\begin{equation}}
\newcommand{\eeq}{\end{equation}}
\newcommand{\beqn}{\begin{eqnarray}}
\newcommand{\eeqn}{\end{eqnarray}}

\usepackage{hyperref}

\usepackage{amssymb}
\usepackage{amsthm}
\usepackage{framed}
\usepackage{amsmath}
\usepackage{geometry}
\newtheorem{lemma}{Lemma}
\newtheorem{thm}{Theorem}
\newtheorem{cor}{Corollary}
\theoremstyle{definition}
\newtheorem{conj}{Conjecture}

\newcommand{\UU}{\mathrm{U}}
\newcommand{\LU}{\mathrm{LU}}

\begin{document}
\title{Critical drag as a mechanism for resistivity}
\author{Dominic V. Else}
\affiliation{Department of Physics, Harvard University, Cambridge, MA 02138, USA}
\affiliation{Department of Physics, Massachusetts Institute of Technology, Cambridge, MA 02139, USA}
\author{T. Senthil}
\affiliation{Department of Physics, Massachusetts Institute of Technology, Cambridge, MA 02139, USA}
\begin{abstract}
A quantum many-body system with a conserved electric charge can have a DC resistivity that is either exactly zero (implying it supports dissipationless current) or nonzero. Exactly zero resistivity is related to conservation laws that prevent the current from degrading. In this paper, we carefully examine the situations in which such a circumstance can occur. We find that exactly zero resistivity requires either continuous translation symmetry, or an internal symmetry that has a certain kind of ``mixed anomaly'' with the electric charge. (The symmetry could be a generalized global symmetry associated with the emergence of unbreakable loop or higher dimensional excitations.) However, even if one of these is satisfied, we show that there is still a mechanism to get nonzero resistivity, through critical fluctuations that drive the susceptibility of the conserved quantity to infinity; we call this mechanism ``critical drag''. Critical drag is thus a mechanism for resistivity that, unlike conventional mechanisms, is unrelated to broken symmetries. We furthermore argue that an emergent symmetry that has the appropriate mixed anomaly with electric charge is in fact an inevitable consequence of compressibility in systems with lattice translation symmetry. Critical drag therefore seems to be the \emph{only} way (other than through irrelevant perturbations breaking the emergent symmetry, that disappear at the renormalization group fixed point) to get nonzero resistivity in such systems. Finally, we present a very simple and concrete model -- the `Quantum Lifshitz Model' -- that illustrates the critical drag mechanism as well as the other considerations of the paper. 

\end{abstract}

\maketitle

\section{Introduction}

This paper is concerned with the issue of electrical resistivity in a system with a conserved charge.
In systems described by Landau Fermi liquid theory, resistivity must arise from quasiparticle scattering, due to impurities, umklapp, or electron-phonon interactions. In this paper, however, we want to make general statements that will hold even beyond Fermi liquid theory.

The opposite of nonzero resistivity is dissipationless current. In a system \emph{without} resistivity, the current can flow freely without degrading, even in the absence of an electric field; that is, there exists an equilibrium state of the system with nonzero expectation value of the current. On the face of it, such a situation is precisely the one that is ruled out by Bloch's theorem \cite{Bohm__1949}, which states that the expectation value of the current is zero in thermal equilibrium. A modern and very general argument for Bloch's theorem has been given in Ref.~\cite{Yamamoto_1502} in the continuum and Ref.~\cite{Watanabe_1904} for lattice models. Nevertheless, it is well known that are systems that exhibit dissipationless current, most famously superfluids, but also an electron gas with an exact continuous translation symmetry. 

In general, a powerful way to think about resistivity, or lack thereof, is in terms of the symmetries of the system, or equivalently the conserved quantities.  The correct way to interpret Bloch's theorem, as we will review here, is that it shows that there will be no dissipationless current, \emph{provided} that there is no conserved quantity that inhibits the current from relaxing. For example, in an electron gas with exact continuous translation symmetry, there is such a conserved quantity, namely the momentum. More subtly, the dissipationless current in a superfluid is protected by a conserved vorticity. Indeed, the idea that nonzero resistivity should be traced back to the absence of conserved quantities that can protect the current has been long been prevalent \cite{Mazur__1969, Suzuki__1971,GurAri_1605}, though here we reinterpret this result by phrasing it in terms of a specific loophole in Bloch's theorem.
This is certainly the origin of nonzero resistivity in familiar examples such as Fermi liquid theory, where the resistivity is caused by scattering processes that do not respect momentum conservation.

By contrast, in a recent paper we proposed \cite{Else_2010}  a mechanism  by which nonzero resistivity can occur \emph{even when} a conserved quantity is present that would   \emph{a priori} permit a circumvention of Bloch's theorem.  The resistivity instead arises from critical fluctuations. (If time reversal and inversion symmetry are present, the operator that is critically fluctuating must be odd under these symmetries.) We will refer to this phenomenon as \emph{\TheTerm}. Critical drag represents a totally different mechanism for resistivity compared with the usual mechanism of breaking the symmetries that would protect the current.

In this work, we will give a more general and self-contained discussion of the concept of \TheTerm.
We will also exhibit a solvable field theory that exhibits \TheTerm.  This field theory -- known as the Quantum Lifshitz Model (QLM) -- describes the phase transition between an ordinary superfluid phase and a different superfluid phase associated with bose condensation at non-zero momentum. The QLM has been studied extensively in a number of different theoretical contexts\cite{ardonne2004topological,vishwanath2004quantum,fradkin2004bipartite,po2015two}. Recently it was proposed\cite{lake2021re} to describe the re-entrant superconductivity observed in twisted trilayer graphene. The gapless excitations of the QLM fixed point are described by a free field theory. This enables explicit calculation of many of its physical properties. Here we will see that it provides a concrete illustration of many of the general considerations of this and our previous papers.

\subsection{Critical drag vs. broken symmetries as mechanisms for resistivity}

A further point that we wish to explore in this work is under which circumstances critical drag is necessary to obtain nonzero resistivity, as opposed to the more familiar mechanism of simply breaking those symmetries that would otherwise protect the current. First one needs to consider under which circumstances a conserved quantity can prevent the current from relaxing. Conventionally \cite{Kadanoff__1963,Lucas_1502} this is expressed in terms of a cross-susceptibility (often referred to as an ``overlap'') between the current and the conserved quantity; in this paper we will point out that this cross-susceptibility  in fact precisely reflects a so-called ``mixed 't Hooft anomaly'' \cite{tHooftAnomaly, Kapustin_1403_0617} between the electric charge $\mathrm{U}(1)$ symmetry and the symmetry generated by the conserved quantity. One consequence of this is that  microscopic internal symmetries can never prevent the current from relaxing; a symmetry that does protect the current must either be a spatial symmetry such as translation symmetry, or else it must be an emergent symmetry not present at the microscopic scale.

Another consequence of the anomaly perspective is that we can make contact with the theory of compressibility developed in Ref.~\cite{Else_2007}. A system is called ``compressible'' if the microscopic electric charge density can be continuously tuned (possibly with other parameters of the Hamiltonian tuned simultaneously) without qualitatively changing the resulting low-energy physics. We will give general arguments (though not a completely rigorous proof) suggesting that a system with microscopic lattice translation symmetry is compressible if and only if it has an emergent symmetry that has a certain kind of mixed 't Hooft anomaly with the electric charge $\mathrm{U}(1)$.  Given the discussion of the previous paragraph, this seems to suggest that, in the absence of critical drag, a clean compressible system always supports dissipationless current, at least in an emergent sense; more precisely, since the precise meaning of an an ``emergent symmetry'' is that it is an exact symmetry of the RG fixed-point theory that controls the low-temperature behavior, it follows that this fixed-point theory must have exactly zero resistivity. (For a real system at finite temperature, there can be irrelevant terms that would not be present in the fixed-point theory, and these can restore nonzero resistivity). This is indeed what occurs in familiar examples of compressible systems without critical drag such as Fermi liquid metals and superfluids.

However, as pointed out in Ref.~\cite{Else_2010}, these considerations present a conundrum if one seeks to explain the $T$-linear resistivity seen in many non-Fermi liquid metals; in many such materials, from looking at how the conductivity scales as a function of frequency and temperature, one can reach the conclusion that the resistivity must be nonzero even in the fixed-point theory. It therefore follows that critical drag must be present in the fixed-point theory controlling these metals. This was basically the argument made in Ref.~\cite{Else_2010}; in this paper we state the argument in a more general and systematic way.

\subsection{Generalities on emergent symmetries and an  illustrative solvable model}
 
 Beyond the general considerations of resistivity and critical drag discussed above, we will also discuss the QLM mentioned above as an interesting addendum to the general theory of compressible systems that we introduced in Ref.~\cite{Else_2010}. In that paper, we studied in a very general way constraints on the low energy physics of compressible quantum phases/phase transitions in systems with a global $\UU(1)$ and (lattice) translation symmetries.   The global $\UU(1)$ symmetry corresponds to conservation of particle number in the microscopic system.   
In general, the renormalization group Infra-Red (IR) fixed point that controls the low energy physics may have a different symmetry than the microscopic symmetry $G_{\mathrm{UV}} $.   Further $G_{\mathrm{UV}}$ will embed into as the emergent symmetry of the IR fixed point as an {\em internal} symmetry. Let the emergent internal symmetry of the IR theory be denoted  $G_{\mathrm{IR}}$. 
In a compressible state, our previous work showed that  $G_{\mathrm{IR}}$ is severely constrained. 

In general the the emergent symmetry $G_{\mathrm{IR}}$ may include both ordinary $0$-form symmetries, as well as what are known as  ``higher-form symmetries"\cite{Gaiotto_1412}.  In the condensed matter context, these higher form symmetries are typically associated with  the emergence of  fractionalized excitations and associated deconfined gauge fields.  In our earlier work we restricted attention to situations where $G_{\mathrm{IR}}$ included either a finite higher-form symmetry, or was simply an ordinary $0$-form symmetry group. This class includes  almost all the known examples of compressible quantum matter. In that case we proved that the $0$-form symmetry included in $G_{\mathrm{IR}}$ is necessarily not a compact finite-dimensional Lie group\footnote{The classic example of a compressible state of matter that does not spontaneously break $G_{\mathrm{UV}}$ is  a Landau Fermi liquid. This has an infinite dimensional internal symmetry associated with conservation of Landau quasiparticles at each point of the Fermi surface. }.  The question of whether a ground state that does not spontaneously break $G_{\mathrm{UV}}$ could have a $G_{\mathrm{IR}}$ that includes a continuous higher-form symmetry was left open, as was the question of what the `filling' constraints on such a state would be.

In this paper we provide a very simple example of a compressible ground state of bosons that does not spontaneously break any microscopic symmetries and has an emergent continuous $1$-form symmetry.  The most familiar ground state of bosons at finite density is a superfluid state which spontaneously breaks the global $\UU(1)$ symmetry.  The low energy theory of a superfluid may be formulated in terms of an action for the phase of the condensing boson. This theory has an emergent continuous $1$-form symmetry (denoted $\UU(1)_1$) associated with the conservation of the winding number of the phase around a closed loop in real space.  This $1$-form symmetry is explicitly broken by gapped vortex excitations of the superfluid. However  the superfluid spontaneously breaks the global $\UU(1)$ symmetry, and hence is not a suitable answer to the question of whether such continuous 1-form symmetries can emerge in ground states that do not spontaneously break $G_{\mathrm{UV}}$.

Instead, the state we describe appears at the quantum Lifshitz critical point between the familiar  superfluid ground state where the bosons condense at zero momentum, and a different superfluid state where the bosons condense at non-zero momentum.      At this quantum Lifshitz critical point, the Bose condensation is suppressed so that $G_{\mathrm{UV}}$ is preserved.    However the vortices in the boson phase are nevertheless gapped finite energy excitations.  The theory thus has an emergent  continuous $\UU(1)_1$ $1$-form symmetry.  Furthermore this critical point is compressible; the boson density can be tuned continuously along the phase boundary.  Despite this, the zero temperature transport is that of an  {\em insulator}. Thus we have a rare example of a compressible insulator at $T= 0$.  The suppression of the conductivity at the fixed point theory can be understood as coming from the critical drag mechanism.  We also discuss how these results are modified by nonzero
temperature.

Ref.~\cite{lake2021bose} discussed  a class of compressible states of interacting bosons at a non-zero density in spatial dimension $d>1$, dubbed `Bose-Luttinger Liquids'.  These Bose-Luttinger Liquids  have a low energy description in terms of gapless phase fluctuations living at a surface in momentum space. They also preserve the full microscopic symmetry $G_{\mathrm{UV}}$ (including in particular the microscopic global $\UU(1)$ associated with particle number conservation). Vortices in the boson phase are gapped excitations, and the low energy phase-only theory has an emergent  higher form symmetry.  This was  suggested to be a  a one-form symmetry in Ref. ~\cite{lake2021bose} but we will show here that there is actually an emergent continuous two-form symmetry. The full set of emergent symmetries have a mixed anomaly with a structure that leads to a metal with zero resistivity.

\subsection{Outline}
The outline of the remainder of the paper is as follows. In Section \ref{sec:bloch}, we discuss the connection between conserved quantities and violations of Bloch's theorem. We show that conserved quantities lead to a loophole in Bloch's theorem \emph{provided} that a certain non-trivial transformation property of the conserved quantity under large gauge transformations is satisfied. We argue that a microscopic internal symmetry can never satisfy such a non-trivial transformation property. In fact, for internal symmetries this transformation property precisely reflects a so-called ``mixed 't Hooft anomaly'' between the  symmetry generated by conserved quantity and the electric charge $\mathrm{U}(1)$. Such an anomaly can, however, occur if the symmetry is emergent. In Sections \ref{sec:emergent_1d} and \ref{sec:emergent_2d}, we give discuss emergent symmetries that appear in familiar examples, such as Luttinger liquids, Fermi liquids, and superfluids, and have the requisite mixed 't Hooft anomaly to lead to a loophole in Bloch's theorem.

In Section \ref{sec:compressibility}, we consider the general relations between the property of having a mixed 't Hooft anomaly that leads to a loophole in Bloch's theorem (which we call \emph{fluxibility}), and compressibility (ability of the system to have a continuously tunable charge filling). We give arguments suggesting that in fact fluxibility and compressibility are equivalent in systems with a microscopic lattice translation symmetry. This implies that the only way to get nonzero resistivity in compressible systems (other than through irrelevant terms that break the emergent symmetry) is through the mechanism of critical drag. We describe this mechanism in Section \ref{sec:critical_drag}.

In Section \ref{sec:qlm}, we consider the QLM as a solvable example of the general considerations of this paper, and in particular as a prototypical example of critical drag. We derive various physical properties of the QLM and relate them to the general considerations.

Finally, in Section \ref{sec:discussion} we conclude and discuss future directions.
Several appendices contain additional details. In particular the Bose-Luttinger Liquids are discussed in Appendix \ref{appendix:BLL}. 

\section{Conservation laws and dissipationless current: a loophole in Bloch's theorem}
\label{sec:bloch}
In this section we expound on the connections between conservation laws and dissipationless current. Suppose we have some system in which the only conserved quantities are the total energy and the total charge $\hat{Q}$ of a global $\mathrm{U}(1)$ symmetry. For ease of exposition, throughout this paper we refer to $\hat{Q}$ as the ``electric charge'', and its corresponding current the ``electric current''. This should not be taken to imply that particles charged under $\hat{Q}$ necessarily are charged under the actual electromagnetic field of the universe.

Then we know that the thermal equilibrium state of the system will be described by the grand canonical ensemble
\begin{equation}
\label{eq:grand_canonical}
\rho = \frac{1}{\mathcal{Z}} \exp\left(-\beta[\hat{H} - \mu \hat{Q}]\right),
\end{equation}
where $\beta$ is the inverse temperature and $\mu$ is the chemical potential. Bloch's theorem amounts to the statement that the expectation value of the current in a state of the form \eqnref{eq:grand_canonical} is zero.

However, the situation is modified if in addition, we also have another conserved quantity $\hat{\Gamma}$ that commutes with $\hat{Q}$. Then the thermal equilibrium state of the system is instead described by a generalized Gibbs ensemble
\begin{equation}
\label{eq:generalized_gibbs}
\rho = \frac{1}{\mathcal{Z}} \exp\left(-\beta[\hat{H} - \mu \hat{Q} - \eta \hat{\Gamma}]\right),
\end{equation}
where we have introduced the additional thermodynamic parameter $\eta$.
The crucial point is that a state of the form \eqnref{eq:generalized_gibbs} does \emph{not} need to have zero expectation value of the electric current, under certain circumstances that we will spell out. In such circumstances, the electric current is prevented from relaxing due to its overlap with the conserved quantity $\hat{\Gamma}$.

In order to show this, let us review the proof of Bloch's theorem in Refs.~\cite{Yamamoto_1502,Watanabe_1904} and show how it can fail in the presence of the additional conserved quantity $\hat{\Gamma}$. For simplicity we first consider a one-dimensional system. 

The grand canonical ensemble state \eqnref{eq:grand_canonical} has the property that
\begin{equation}
\label{eq:VKV}
\langle V^{\dagger} \hat{K} V \rangle \geq \langle \hat{K} \rangle,
\end{equation}
for any unitary $V$, where $\langle \cdot \rangle$ denotes expectation values with respect to $\rho$, and where we have defined
\begin{equation}
\label{eq:Kamiltonian}
 \hat{K} = \hat{H} - \mu \hat{Q}
 \end{equation}
This is a special case of the general statement that the state $\rho$  minimizes the grand potential
\begin{equation}
\Phi = \langle \hat{K} \rangle - T S(\rho),
\end{equation}
where $T = \beta^{-1}$ and $S(\rho) = -\mathrm{Tr} (\rho \ln \rho)$ is the von Neumann entropy. We obtain \eqnref{eq:VKV} by noting that $ V \rho V^{\dagger}$ has the same von Neumann entropy as $\rho$ itself. 

Now we define the gauge transformation operator $U_\lambda$ according to
\begin{equation}
U_\lambda = \exp\left(-i \int \lambda(x) \hat{n}(x) dx \right),
\end{equation}
where $\hat{n}(x)$ is the local electric charge density, such that $\hat{Q} = \int \hat{n}(x) dx$.
The main technical result required to prove Bloch's theorem is that, if $\lambda(x)$ is slowly varying, then
\begin{equation}
\label{eq:UHU}
U_\lambda \hat{H} U_\lambda^{\dagger} = \hat{H} + \int  [\partial_x \lambda(x) ] \hat{j}(x) dx + \cdots,
\end{equation}
where $\hat{j}(x)$ is the operator measuring the electric current at the point $x$, and the terms contained in the ``$\cdots$'' involve higher powers of $\partial_x \lambda$ and/or higher derivatives.
If we take the expectation value of the right hand side (not including the higher-order terms), we obtain
\begin{equation}
\langle \hat{H} \rangle + 2\pi j w_\lambda,
\end{equation}
where $w = \frac{1}{2\pi}\int \partial_x \lambda(x) dx$ is the winding number of $\lambda$ (assuming periodic boundary conditions), and $j := \langle \hat{j}(x) \rangle$ (which must be independent of $x$). The precise statement proven in Ref.~\cite{Watanabe_1904} is that if we consider the system with periodic boundary conditions, on a ring of length $L$, and take $\lambda(x) = \pm 2\pi x/L$ (giving $w_\lambda = \pm 1$), then
\begin{equation}
\langle U_\lambda H U_\lambda^{\dagger} \rangle = \langle \hat{H} \rangle \pm 2\pi j + O\left(\frac{1}{L}\right).
\end{equation}
Combining with \eqnref{eq:VKV} (and noting that $U_\lambda \hat{Q} U_\lambda^{-1} = \hat{Q}$), we then conclude that
\begin{equation}
\label{eq:bloch_theorem}
|j| = O\left( \frac{1}{L} \right),
\end{equation}
and so $j$ goes to zero in the thermodynamic limit $L \to \infty$, which is Bloch's theorem.

Now we are in a position to identify what goes wrong if we try to apply this argument to the generalized Gibbs ensemble \eqnref{eq:generalized_gibbs}. First we should define
\begin{equation}
\hat{K} = \hat{H} - \mu \hat{Q} - \eta \hat{\Gamma}
\end{equation}
instead of \eqnref{eq:Kamiltonian}. Then the argument proceeds more or less as before, except that it is not necessarily the case that $U_\lambda \hat{\Gamma} U_\lambda^{-1} = \hat{\Gamma}$. Let us suppose that
\begin{equation}
\label{eq:Xi}
\langle U_\lambda \hat{\Gamma} U_\lambda^{-1} \rangle = \langle \hat{\Gamma} \rangle + \int \langle \hat{\Xi}(x) \rangle \partial_x \lambda(x) dx + \cdots
\end{equation}
for some local operators $\hat{\Xi}(x)$, and where the condition on the terms in the ``$\cdots$'' is that they should contribute at most $O(\frac{1}{L})$ once we set $\lambda(x) = \pm 2\pi x /L$ and take expectation values.
The reason why \eqnref{eq:Xi} only involves $\partial_x \lambda$ and not $\lambda$ itself is that if $\lambda(x)= \lambda$ is independent of $x$, then $U_\lambda = e^{i\lambda \hat{Q}}$, and by assumption $\hat{Q}$ commutes with $\hat{\Gamma}$.

 Then, repeating the above argument, we find instead of \eqnref{eq:bloch_theorem} that
\begin{equation}
\label{eq:j_in_terms_of_eta}
| j - \eta \Xi | =  O\left(\frac{1}{L}\right).
\end{equation}
where
\begin{equation}
\Xi = \frac{1}{L} \int \langle \hat{\Xi}(x) \rangle dx
\end{equation}
Thus, in the thermodynamic limit we have
\begin{equation}
\label{eq:j_eta_Xi}
j = \eta \Xi
\end{equation}
 rather than zero. Thus, we have found a loophole in Bloch's theorem, in the presence of additional conserved quantities.

Let us further discuss the significance of the condition $\Xi \neq 0$ that is required in order to have dissipationless electric current. First we want to show that if $\hat{\Gamma}$ is the generator of an internal symmetry at the lattice scale, then one would expect that $\Xi = 0$. To see this, note that in this case one would expect the symmetry to be ``on-site'', in the sense that one can write
\begin{equation}
\label{eq:M_on_site}
\hat{\Gamma} = \sum_i \hat{\gamma}^{(i)},
\end{equation}
where the sum is over lattice sites $i$, and the operator $\hat{\gamma}^{(i)}$ acts only on the degrees of freedom on site $i$.
Similarly the electric charge (which we assume always generates a microscopic symmetry) can be written as
\begin{equation}
\hat{Q} = \sum_i \hat{q}^{(i)}
\end{equation}
where $\hat{q}^{(i)}$ is the electric charge on site $i$. Since, by assumption, $\hat{\Gamma}$ commutes with $\hat{Q}$, it follows that $[\hat{\gamma}^{(i)}, \hat{q}^{(i)}] = 0$ on each site $i$. Defining a gauge transformation operator at the lattice scale according to
\begin{equation}
U_\lambda = \exp\left(-i \sum_i \lambda_i \hat{q}^{(i)}\right),
\end{equation}
we then immediately find that $U_\lambda \hat{\Gamma} U_\lambda^{-1} = \hat{\Gamma}$, which means that $\hat{\Xi}(x) = 0$ in \eqnref{eq:Xi}.

On the other hand, in general it is possible to have $\Xi \neq 0$ if $\hat{\Gamma}$ cannot be represented in the on-site form \eqnref{eq:M_on_site}. One possibility is that $\hat{\Gamma}$ represents momentum (not lattice momentum) $\hat{P}$, i.e. it is the generator of a continuous translation symmetry,. Then the generalized Gibbs ensemble \eqnref{eq:generalized_gibbs} takes the form
\begin{equation}
\rho = \frac{1}{\mathcal{Z}} \exp\left(-\beta[\hat{H} - \mu \hat{Q} - v\hat{P}]\right)
\end{equation}
where the parameter $v$, as we shall see shortly, can be interpreted as the overall drift velocity of the system. One can then show that
\begin{equation}
U_\lambda \hat{P} U_\lambda^{\dagger} = \hat{P} + \int \hat{n}(x) \partial_x \lambda(x) + \cdots,
\end{equation}
where $\hat{n}(x)$ is the electric charge density of the system.
Thus, the operator $\hat{\Xi}(x)$ in \eqnref{eq:Xi} is $\hat{n}(x)$, and from \eqnref{eq:j_eta_Xi} we find that in the thermodynamic limit, the electric current is given by
\begin{equation}
 j = n v,
\end{equation}
  where $n = \langle {\hat{n}(x)} \rangle$ (which is independent of $x$ given the continuous translation symmetry) is the expectation value for the electric charge density of the system.
  
  Thus, we have reproduced a well-known fact: systems with momentum conservation, and nonzero electric charge density, generally can carry dissipationless electric current. On the other hand, in condensed matter physics we are generally interested in systems on the lattice that do not have such a conservation law. Note that conservation of lattice momentum (as opposed to continuous momentum) is not sufficient on its own to obtain dissipationless electric current, as current can degrade by umklapp scattering. Therefore, in this work we will not be too interested in dissipationless current protected by momentum conservation. Instead, in the following sections we will instead consider dissipationless current protected by an \emph{emergent} symmetry that is internal but nevertheless cannot be written in the form \eqnref{eq:M_on_site}.

\section{Emergent symmetries and anomalies in 1D}
\label{sec:emergent_1d}
Let us suppose that $\hat{\Gamma}$ generates an \emph{emergent} symmetry that is a symmetry of the effective theory governing the dynamics of the system at long length scales, but not of the original microscopic Hamiltonian. In general such symmetries cannot be written in the form \eqnref{eq:M_on_site}, and therefore have the potential to protect dissipationless electric current.

In fact, for internal symmetries, $U_\lambda \hat{\Gamma} U_\lambda^{-1} \neq \hat{\Gamma}$, which is the condition required to protect dissipationless electric current as we saw above, is a signature of something called a `` 't Hooft anomaly''  \cite{tHooftAnomaly, Kapustin_1403_0617}, which is a common property of emergent symmetries. A symmetry is said to have a 't Hooft anomaly if the conserved quantity corresponding to the symmetry becomes non-conserved in the presence of a background gauge field. In the present case, we are considering mixed anomalies between the symmetry $\hat{\Gamma}$ and the electric charge $\hat{Q}$, so a mixed anomaly will occur when $\hat{\Gamma}$ is nonconserved in response to an applied electromagnetic field (i.e.\ a background gauge field for the symmetry generated by $\hat{Q}$), as described by the ``anomaly equation'', which is an operator equation
\begin{equation}
\label{eq:0form_anomaly}
\partial_t \hat{n}_\Gamma(x,t) + \partial_x \hat{j}_\Gamma(x,t) = \frac{m}{2\pi} E(x,t)
\end{equation}
where $\hat{n}_\Gamma$ is the local density of $\hat{\Gamma}$ charge, $\hat{j}_\Gamma$ is the current of $\hat{\Gamma}$ charge, $E(x,t)$ is the electric field at a given time and position, and the constant $m$ is the anomaly coefficient. Here time dependence corresponds to Heisenberg evolution of operators.

In particular one can show (see Appendix \ref{appendix:commutator}) that this 't Hooft anomaly implies that the local densities $\hat{n}(x)$ and $\hat{n}_\Gamma(x)$ (of $\hat{Q}$ and $\hat{\Gamma}$ charges respectively) fail to commute; instead they have a non-trivial commutator
\begin{equation}
\label{eq:rho_rho_M}
[\hat{n}(x), \hat{n}_\Gamma(x')] = -\frac{i m}{2\pi} \delta'(x - x')
\end{equation}
where $\delta'$ is the derivative of the Dirac delta function.
From \eqnref{eq:rho_rho_M}, one can derive \eqnref{eq:Xi} with $\hat{\Xi}(x) = m/(2\pi)$. So in the thermodynamic limit, the current is given by
\begin{equation}
\label{eq:j_anomaly_1d}
j = \frac{m}{2\pi} \eta,
\end{equation}
where $\eta$ is the thermodynamic variable conjugate to $\hat{M}$.

\subsection{Example: Luttinger liquid in 1D}
As an example, consider a Luttinger liquid in 1D. A well-known fact about such a system is that, in the low-energy effective theory, there are separately conserved left- and right-moving charges $\hat{Q}_L$ and $\hat{Q}_R$. Another way to say this is that in addition to the total electric charge $\hat{Q} = \hat{Q}_L + \hat{Q}_R$, there is another conserved quantity, the axial charge $\hat{Q}_A = \hat{Q}_L - \hat{Q}_R$.

The axial charge has a mixed anomaly with the total electric charge. Indeed, if one applies an electric field $E$, one can show that the continuity equation for the axial charge is violated by a term $E/\pi$ on the right hand side.  Comparing with \eqnref{eq:0form_anomaly}, we see that there is a mixed anomaly with coefficient $m = 2$. Therefore, by the general discussion above, the low-energy effective theory of a Luttinger liquid can carry dissipationless electric current. A related connection between the chiral anomaly and dissipationless current was also noted for chiral Luttinger liquids (such as those occurring at the boundary of a quantum hall system) in Ref.~ \cite{kapustin2019absence}, where the electric charge itself has an anomaly.

Of course, in the real microscopic system, at any nonzero temperature, one expects the current to be degraded due to umklapp scattering; however for incommensurate filling, such scattering is irrelevant in the RG sense and hence not present in the RG fixed point theory. (This is an example of a case where interactions that are 	``irrelevant'' in the RG sense are still important for determining the leading low-temperature scaling of physical quantities such as the current decay rate. Such interactions are generally referred to as ``dangerously irrelevant'').

\section{Emergent symmetries and anomalies in 2D: 1-form and loop group symmetries}
\label{sec:emergent_2d}
In this section we discuss generalizations of the considerations of Section \ref{sec:emergent_1d} to 2 spatial dimensions. The same basic principles will apply: the condition for dissipationless electric current is that there should be an emergent symmetry that has a mixed anomaly with the electric charge. However, the precise details of the symmetry and anomaly will be different. There are two main classes of symmetries we will want to consider: 1-form symmetries and loop-group symmetries. The anomaly structure was previously discussed in Ref.~\cite{Delacretaz_1908} for 1-form symmetries and in Ref.~\cite{Else_2007} for loop-group symmetries. Let us also mention that the emergent symmetry group for the Bose-Luttinger liquid of Ref.~\cite{lake2021bose} is more exotic than the ones considered here but also gives rise to dissipationless current; we discuss the details in Appendix \ref{appendix:BLL}.

\subsection{1-form symmetries}
\label{subsec:1form}
The concept of a 1-form symmetry originated in Ref.~\cite{Gaiotto_1412}. For the case of continuous 1-form symmetries (which we will mostly be concerned with in the present paper), they can most straightforwardly be introduced as a special kind of local conservation law. Recall that local conservation of (0-form) charge amounts to the statement that there exists a space-time vector $J^\mu$ such that the continuity equation is satisfied:
\begin{equation}
\label{eq:continuity}
\partial_\mu J^\mu = 0.
\end{equation}
(Here and below we use the implicit index summation convention.) The time component $n = J^0$ is interpreted as the charge density, and \eqnref{eq:continuity} implies that the integral
\begin{equation}
Q = \int  n(\mathbf{x}) d^d x
\end{equation}
over a space without boundary is independent of time, i.e.\ it is conserved.

Now we define local conservation of 1-form charge to be the statement that there exists an \emph{anti-symmetric tensor} $\mathcal{J}^{\mu \nu}$ such that
\begin{equation}
\label{eq:local_cons_1form}
\partial_\mu \mathcal{J}^{\mu \nu} = 0.
\end{equation}
We define the ``charge density'' $u^i = \mathcal{J}^{0 i}$ (which is a spatial vector).
If we now let $\Sigma$ be a closed $(d-1)$-dimensional surface in $d$-dimensional space, then we can define the associated charge $\mathcal{Q}_\Sigma$ by the surface integral
\begin{equation}
\mathcal{Q}_\Sigma = \int_\Sigma \mathbf{u} \cdot d\boldsymbol{\Sigma},
\end{equation}
where $d\boldsymbol{\Sigma}$ is the infinitesimal vector normal to the surface. \eqnref{eq:local_cons_1form} implies that $\mathcal{Q}_\Sigma$ is conserved. Note, however, that \eqnref{eq:local_cons_1form} also implies that $\mathcal{Q}_\Sigma$ is identically zero for any surface $\Sigma$ with an interior. On topologically non-trivial manifolds such a torus, it is possible to have closed surfaces without an interior, for which $\mathcal{Q}_\Sigma$ can be a non-trivial quantity. (Note, however, that the \emph{local} charge conservation \eqnref{eq:local_cons_1form} is still a  non-trivial constraint without needing to worry about about the global topology of space, hence why 1-form symmetries have non-trivial consequences for the local physics regardless of the boundary conditions.)

A well-known example of a system with a 1-form symmetry is electromagnetism in 3 spatial dimensions (without magnetic monopoles), for which one defines $\mathcal{J}^{\mu \nu} = \epsilon^{\mu \nu \lambda \sigma} F_{\lambda \sigma}$, where $F_{\lambda \sigma}$ is the field-strength tensor;  \eqnref{eq:local_cons_1form} then follows from Maxwell's equations. In this case, the 1-form charge density is the magnetic field, and the charge $\mathcal{Q}_\Sigma$ is the magnetic flux through the surface $\Sigma$. 

However, another example that will be more relevant for the current paper is a superfluid in 2 spatial dimensions. In this case, we can define
\begin{equation}
\label{eq:J1form_superfluid}
\mathcal{J}^{\mu \nu} = \frac{1}{2\pi} \epsilon^{\mu \nu \lambda} \partial_\lambda \theta,
\end{equation}
where $\theta$ is the superfluid phase field. Below the superfluid transition temperature, there are no vortices in the system, so $(\partial_\mu \partial_\nu - \partial_\nu \partial_\mu) \theta = 0$, which immediately implies \eqnref{eq:local_cons_1form}. The charge $\mathcal{Q}_\Sigma$ associated with a closed 1-dimensional surface $\Sigma$ (that is, a closed loop) is the winding number of the superfluid phase over the loop. More generally a superfluid in $d$ spatial dimensions will have a $(d-1)$-form symmetry.

1-form symmetries in 2D can also give rise to a loophole in Bloch's theorem in a similar way to what we saw above for 0-form symmetries in 1D. Firstly, we note that the analog for 1-form symmetries of the generalized Gibbs ensemble \eqnref{eq:generalized_gibbs} above is
\begin{equation}
\label{eq:1_form_gibbs}
\rho = \frac{1}{\mathcal{Z}} \exp\left(-\beta\left[\hat{H} - \mu \hat{Q} - \int \eta_i(\mathbf{x}) \hat{u}^{i}(\mathbf{x}) d^d \mathbf{x}\right]\right),
\end{equation}
where $\eta_i$ is a closed 1-form, i.e.\ a vector field on space such that $\partial_i \eta_j - \partial_j \eta_i = 0$. One can justify this ensemble using the maximum entropy  principle of statistical mechanics: the state \eqnref{eq:1_form_gibbs} maximizes the von Neumann entropy subject to the constraints that the expectation values of $\hat{H}$, the electric charge $\hat{Q}$, and the 1-form charge $\hat{\mathcal{Q}}_\Sigma$ on any closed loop $\Sigma$ are held fixed.  We show this in Appendix \ref{appendix:generalized_gibbs}.

A mixed 't Hooft anomaly between the electric charge and the 1-form symmetry means that firstly, the local charge conservation of 1-form charge is violated in the presence of an electromagnetic field $F_{\mu \nu}$ (i.e. a background gauge field for the electric charge):
\begin{equation}
\label{eq:nonconservation_1form}
\partial_\mu \mathcal{J}^{\mu \nu} = \frac{m}{4\pi} \epsilon^{\nu \lambda \sigma} F_{\lambda \sigma},
\end{equation}
where, if the 1-form charge is quantized to be an integer (as it is in the examples discussed above), the anomaly coefficient $m$ is quantized to be an integer. From this one can show, similarly to the 1-D case above, that the local densities fail to commute:
\begin{equation}
[\hat{n}(\mathbf{x}), \hat{u}^{i}(\mathbf{x}')] = -i \frac{m}{2\pi} \epsilon^{ij} \partial_{x_j} \delta^2(\mathbf{x} - \mathbf{x}').
\end{equation}
In turn this implies that
\begin{equation}
U_\lambda \mathcal{Q}_\eta U_\lambda^{-1} = \mathcal{Q}_\eta +  \frac{m}{2\pi} \int \epsilon^{ij} \eta_i(\mathbf{x}) \partial_j \lambda(\mathbf{x}) d^2 \mathbf{x},
\end{equation}
where we defined
\begin{align}
\mathcal{Q}_\eta &= \int \eta_i(\mathbf{x}) u^i(\mathbf{x}) d^2 \mathbf{x} \\
U_\lambda &= \exp\left( -i \int \lambda(\mathbf{x}) \hat{n}(\mathbf{x}) d^2 \mathbf{x} \right).
\end{align}
We can also straightforwardly generalize the above results to $(d-1)$-form symmetries in $d$ spatial dimensions, in which case $\hat{u}$ becomes a rank $(d-1)$ antisymmetric contravariant tensor, and $\eta$ becomes a closed $(d-1)$-form.

 In general space dimension $d$,
if we repeat the argument for Bloch's theorem (see Appendix \ref{appendix:bloch_general_dimension}), one therefore concludes that the electric current flowing through any closed $(d-1)$-dimensional surface $\Sigma$ satisfies
\begin{equation}
\label{eq:1form_bloch_theorem}
I_\Sigma := \int_\Sigma \mathbf{j}(\mathbf{x}) \cdot d\boldsymbol{\Sigma} = \frac{m}{2\pi} \int_\Sigma \eta + O(\Gamma_\Sigma),
\end{equation}
where $\Gamma_\Sigma$ is a geometrical factor with units of $(\mathrm{length})^{d-2}$, and $\int_\Sigma \eta$ denotes the integral of the differential form $\eta$ over $\Sigma$. Unlike in the 1D case, the $O(\Gamma_\Gamma)$ term (as in the argument for Bloch's theorem in higher dimensions in the absence of anomalies \cite{Watanabe_1904}) does not necessarily go to zero in the thermodynamic limit, however its contribution to the average current density $I_\Sigma/|\Sigma|$ [where $|\Sigma|$ is the $(d-1)$-dimensional area of $\Sigma$] does go to zero as long as all the dimensions of the system go to infinity; see Ref.~\cite{Watanabe_1904} and Appendix \ref{appendix:bloch_general_dimension} for details.

Another way to write \eqnref{eq:1form_bloch_theorem}, in the case $d=2$, is that the electric current density  satisfies
\begin{equation}
\label{eq:jcong}
j^i(\mathbf{x}) \cong \frac{m}{2\pi} \epsilon^{ij}  \eta_j(\mathbf{x}) + \cdots,
\end{equation}
where we have introduced the notation that the 	``$\cong$'' sign indicates that equality need hold only when the current density is integrated over any closed $(d-1)$-dimensional surface to find the current flowing through the surface\footnote{Thus, we are explicitly ignoring circulating contributions to the local current density, that do not contribute to the current flowing through any closed surface.}, and the ``$\cdots$'' refers to the corrections that give rise to the $O(\Gamma_\Sigma)$ term in \eqnref{eq:1form_bloch_theorem}.

Thus, we have shown that the existence of a $(d-1)$-form symmetry that has a mixed anomaly with the electric charge ensures a loophole to Bloch's theorem in $d$ spatial dimensions.

\subsubsection{Example: 2D superfluid}
A simple example of the above is a superfluid in 2D. As in \eqnref{eq:J1form_superfluid} above, we define the 1-form current density as
\begin{equation}
\label{eq:J1form_superfluid_gaugeinvariant}
\mathcal{J}^{\mu \nu} = \frac{1}{2\pi} \epsilon^{\mu \nu \lambda} (\partial_\lambda \theta - A_\lambda),
\end{equation}
where here we have taken into account the possibility of a background gauge field $A_\lambda$ for the electric $\mathrm{U}(1)$ 0-form symmetry and written the current in the form \eqnref{eq:J1form_superfluid_gaugeinvariant} to ensure that it is gauge-invariant. Then we immediately find that, in the absence of vortices,
\begin{equation}
\label{sfanom}
\partial_\mu \mathcal{J}^{\mu \nu} = \frac{1}{4\pi} \epsilon^{\nu \lambda \sigma} F_{\lambda \sigma},
\end{equation}
where $F_{\lambda \sigma} = \partial_\lambda A_\sigma - \partial_\sigma A_\lambda$ is the electromagnetic field strength tensor. This agrees with the anomaly equation \eqnref{eq:nonconservation_1form} with $m = 1$.

Thus, the conservation of 1-form charge in a superfluid leads to dissipationless electric current according to the general mechanism described above.  The current-carrying state is seen to be an equilibrium state of the system, once we take into account the conservation law for 1-form charge.

\subsection{Loop group symmetries}
A \emph{loop group} is a kind of infinite-dimensional Lie group. Specifically, for the case of the loop group $\mathrm{LU}(1)$, the elements of the group are smooth maps from the circle into $\mathrm{U}(1)$. We can formally write these elements as
\begin{equation}
\exp\left(-i \int f(\theta) \hat{N}(\theta) d\theta \right)
\end{equation}
where $\theta$ is some coordinate that parameterizes the circle, $f(\theta)$ is some $\mathrm{U}(1)$-valued function on the circle, and $\hat{N}(\theta)$ can be thought of as the generators of the symmetry. (Though strictly speaking, $\hat{N}(\theta)$ has meaning only when integrated over $\theta$ against a smooth function $f(\theta)$; thus, we can think of it as an operator-valued distribution.) Here we use the notation $\hat{N}(\theta)$ instead of the $\hat{n}(\theta)$ of Ref.~\cite{Else_2007}, since in this paper we have elsewhere reserved lower case $\hat{n}$ for a local density at a point $\mathbf{x}$ in space, while $\hat{N}$ is integrated over all space.
The loop group $\mathrm{LU}(1)$ is the emergent symmetry group of a Fermi liquid in 2D, and probably many non-Fermi liquid metals as well \cite{Else_2007}.

We will assume that the electric charge $\mathrm{U}(1)$ symmetry is a subgroup of $\mathrm{LU}(1)$; that is, the total electric charge can be identified as
\begin{equation}
\hat{Q} = \int \hat{N}(\theta) d\theta.
\end{equation}
Furthermore, the microscopic translation symmetries act in the low-energy theory as
\begin{equation}
\mathbb{T}_i = \exp\left(-i\int k_i(\theta) \hat{N}(\theta) d\theta\right)
\end{equation}
where $i$ ranges over $x$,$y$; this
defines the Fermi surface $k_i(\theta)$ in momentum space. The generalized Gibbs ensemble taking into account the loop-group symmetry takes the form
\begin{equation}
\rho = \frac{1}{\mathcal{Z}} \exp\left(-i\left[\hat{H} - \int \mu(\theta) \hat{N}(\theta) d\theta \right] \right),
\end{equation}
where $\mu(\theta)$ are thermodynamic variables.

The 't Hooft anomalies of an $\mathrm{LU}(1)$ symmetry were discussed in Ref.~\cite{Else_2007};
we review the details in Appendix \ref{appendix:loop_group_anomalies}. Here we just state that, upon applying the general Bloch's theorem argument, one finds the electric current density is given by
\begin{equation}
\label{eq:loop_group_current}
j^i =  \frac{m}{(2\pi)^2} \int \epsilon^{ij} [\partial_\theta k_j(\theta)] \mu(\theta) d\theta,
\end{equation} 
where the integer $m$ is the anomaly coefficient.

One can verify in particular that, if we set $m=1$, this agrees with the usual formula for current in a Fermi liquid. To show this we need to find the relation between $\mu(\theta)$ and $n(\theta) := \langle \hat{n}(\theta,\mathbf{x}) \rangle$
 in a Fermi liquid, where $\hat{n}(\theta,\mathbf{x})$ is the local density whose integral over $\mathbf{x}$ gives $\hat{N}(\theta)$. We do this by invoking the fact that
\begin{equation}
\delta n(\theta) = \frac{1}{(2\pi)^2} |\partial_\theta \mathbf{k}(\theta)| \delta k_F(\theta),
\end{equation}
where $\delta k_F(\theta)$ is the perturbation to the Fermi surface compared with the equilibrium state of the system that has $\mu(\theta)$ independent of $\theta$. We impose the requirement that the perturbed Fermi surface must retain the property that creating a quasiparticle exactly at the new Fermi surface leaves the expectation value of $\hat{H} - \int \mu(\theta) \hat{N}(\theta) d\theta$ invariant. Now, the energy cost of creating a quasiparticle at the Fermi surface is
\begin{equation}
\mathcal{E}_F(\theta) + \int V(\theta, \theta') n(\theta') d\theta',
\end{equation}
where $\mathcal{E}_F(\theta)$ is the single-particle energy at the Fermi surface, and $V(\theta, \theta')$ are the Landau interactions at the Fermi surface. Hence, what we want to impose is that
\begin{equation}
\label{eq:quasiparticle_condn}
\delta \mathcal{E}_F(\theta) - \delta \mu(\theta) + \int V(\theta, \theta') \delta n(\theta') d\theta' = 0
\end{equation}
Now using $\delta \mathcal{E}_F(\theta) = v_F(\theta) \delta k_F(\theta)$, where $v_F(\theta)$ is the single-particle Fermi velocity, and recalling that the current in the unperturbed state is zero, by combining \eqnref{eq:loop_group_current} (with $m=1$) and \eqnref{eq:quasiparticle_condn} we find that
\begin{equation}
\label{eq:fermi_liquid_current}
j^i = \int v_F(\theta) \delta n(\theta) \mathbf{\hat{w}}(\theta) d\theta + \int V(\theta, \theta') \delta n(\theta') \mathbf{w}(\theta) d\theta d\theta',
\end{equation}
where we defined the vector $\mathbf{w}(\theta)$ by its components
\begin{equation}
w^i(\theta) = \epsilon^{ij} \partial_\theta k_j(\theta),
\end{equation}
which points normal to the Fermi surface, and the corresponding unit vector $\mathbf{\hat{w}}(\theta) = \mathbf{w}(\theta)/|\mathbf{w}(\theta)|$.
 \eqnref{eq:fermi_liquid_current} is indeed the standard formula for the current in a Fermi liquid.

\section{Relation with compressibility}
\label{sec:compressibility}
As we have seen above, the condition for dissipationless electric current (in the absence of a microscopic continuous translation symmetry) in the low-energy effective theory is that there should be an emergent continuous symmetry (possibly a higher-form symmetry) that has a particular kind of mixed anomaly with the electric charge $\mathrm{U}(1)$. In general, we say that a system is \emph{fluxible} if its emergent symmetry and anomaly structure is such as to lead to a loophole in Bloch's theorem along the lines discussed above.

Another context where emergent symmetry and their mixed anomalies with the electric charge $\mathrm{U}(1)$ are important is in the filling framework developed in Ref.~\cite{Else_2007}. As shown in Ref.~\cite{Else_2007}, for a system with microscopic lattice translation symmetry and electric charge conservation symmetry, the average electric charge per unit cell $\nu$, referred to as the ``filling'', can be computed modulo 1 if one knows the emergent symmetry of the low-energy theory of the system, its anomaly, and the way the microscopic translations embed into the emergent symmetry. We call the system \emph{compressible} if it is possible, by varying the way that the microscopic symmetries embed into the emergent symmetry group\footnote{In principle one could also imagine tuning the filling by tuning the anomaly instead. However, generally one expects anomalies. at least in the sense we use the term here, to be discrete, so that this would not be possible; at any rate this is what we assume here.} to continuously tune the filling $\nu$. Thus, we here consider ``compressibility'' to be a property of the low-energy theory. In terms of the microscopic Hamiltonian, it would mean that it is possible to tune the filling (possibly tuning parameters of the Hamiltonian simultaneously) and still have the emergent physics described by the same low-energy theory.

At this point we wish to formulate the following conjecture:

\begin{framed}

\begin{conj}
 A system with microscopic lattice translation symmetry and electric charge conservation is compressible if and only if it is fluxible.
 \end{conj}

\end{framed}

We emphasize that this conjecture is distinct from the result proved in Ref.~\cite{Else_2007}, that the emergent symmetry group of a compressible system cannot be a compact finite-dimensional Lie group in spatial dimension $d \geq 2$. By contrast to that result, our conjecture is postulated to hold also in $d=1$ and for systems with emergent higher-form symmetries.

As evidence for this conjecture, note that for all of the emergent symmetries considered in Sections \ref{sec:emergent_1d} and \ref{sec:emergent_2d}, the system is fluxible if and only if the the anomaly coefficient is nonzero. Moreover, all of these examples were previously considered in Ref.~\cite{Else_2007} as examples of systems that are compressible (except when the anomaly coefficient is zero). Meanwhile, an example of a system that is not fluxible would be one in which the emergent symmetry group is just $\mathrm{U}(1)$, or more generally $G_{\mathrm{finite}} \times \mathrm{U}(1)$, where $G_{\mathrm{finite}}$ is a finite group (or even a finite 2-group, that can include both 0-form and 1-form symmetries), since it is only additional continuous symmetries that can modify the thermodynamic ensemble from the standard grand canonical ensemble \eqnref{eq:grand_canonical}. However, in the framework of Ref.~\cite{Else_2007} one can show that the filling would then have to be a rational number.
Thus, in this case the system is neither compressible nor fluxible.

As further evidence for the conjecture, in Appendix \ref{appendix:compressibility} we prove it in spatial dimension $d=1$ using the filling framework of Ref.~\cite{Else_2007}, subject to some mild technical assumptions. This result actually provides strong evidence for the conjecture in higher dimensions, since one should be able to treat higher dimensional systems by compactifying all of the dimensions except one and considering them as quasi-one-dimensional.

\section{Suppressing the conductivity through ``\TheTerm''}
\label{sec:critical_drag}
Assuming the conjecture of the previous section holds, it would suggest that compressible systems with lattice translation symmetry and electric charge conservation \emph{always} can carry dissipationless electric current, at least in the IR fixed-point theory.
In turn, this would suggest that the DC resistivity of the IR fixed-point theory should be exactly zero. As we previously discussed in Ref.~\cite{Else_2010}, this presents a conundrum in particular for describing ``strange metals'' that occur in cuprates and heavy-fermion systems, since experimental evidence suggests that if they are controlled by an IR fixed-point theory, then this theory must have have nonzero DC resistivity. Ref.~\cite{Else_2010} discussed a possible loophole through a mechanism involving quantum criticality. In this section, we will review the mechanism proposed in Ref.~\cite{Else_2010} in a more general context. It is this mechanism that we will refer to in the present paper as ``\TheTerm''. 

To illustrate the mechanism, let us consider, as in Section \ref{sec:emergent_1d}, a system with an additional conserved quantity $\hat{\Gamma}$ as well as the electric charge $\hat{Q}$. At time $t = -\infty$, it is in the grand canonical ensemble \eqnref{eq:grand_canonical}. Now consider how the system responds to an electric field impulse
\begin{equation}
\label{eq:E_impulse}
E(t) = \mathcal{E} \delta(t),
\end{equation}
which we will treat in linear response theory. Then the electric current of the system at the system at time $t$ defines the real-time conductivity $\sigma(t)$:
\begin{equation}
j(t) = \sigma(t) \mathcal{E}.
\end{equation}
If $\lim_{t \to \infty} \sigma(t) := \mathcal{W}/\pi \neq 0$, then in frequency space this corresponds to a delta function in $\sigma(\omega)$ at $\omega = 0$,
\begin{equation}
\sigma(\omega) = \mathcal{W} \delta(\omega) + \cdots,
\end{equation}
and in particular we have that the DC conductivity $\sigma(\omega = 0)$ is infinite.

Why would we have $\mathcal{W} \neq 0$? We can imagine that at late times the system thermalizes to a generalized Gibbs ensemble. But since the electric field can violate conservation of $\hat{\Gamma}$ while it is applied (i.e. at $t=0$) -- in fact, as described previously this is precisely one signature of a mixed 't Hooft anomaly between $\hat{\Gamma}$ and $\hat{Q}$ -- it may be that this ensemble has $\eta \neq 0$, which would imply that it carries nonzero electric current according to \eqnref{eq:j_anomaly_1d}.

More precisely, one can show from the anomaly equation \eqnref{eq:0form_anomaly} that the change in expectation value of the density $\hat{n}_\Gamma$ due to the electric field impulse \eqnref{eq:E_impulse} is
\begin{equation}
\delta \langle \hat{n}_\Gamma \rangle = \frac{m}{2\pi} \mathcal{E},
\end{equation}
where $m$ is the anomaly coefficient.
On the other hand we know that
\begin{equation}
\delta \langle \hat{n}_\Gamma \rangle = \chi_{\Gamma\Gamma} \delta \eta,
\end{equation}
where we have defined the susceptibility
\begin{equation}
\chi_{\Gamma\Gamma} = \frac{d}{d\eta} \langle \hat{n}_\Gamma \rangle _{\eta} \biggr|_{\eta = 0},
\end{equation}
where $\langle \cdot \rangle$ denotes expectation values taken with respect to the generalized Gibbs ensemble,
and we have ignored the possibilities of nonzero cross-susceptibilities between $\hat{\Gamma}$ and the electric charge $\hat{Q}$ or the energy $E$; for example, if time reversal symmetry is present it forces these cross-susceptibilities to be zero \footnote{Specifically, $\hat{\Gamma}$ \emph{must} be time-reversal odd if a nonzero anomaly coefficient $m$ is to be consistent with time-reversal symmetry, as can be seen from \eqnref{eq:rho_rho_M} for example}. (In any case, the main conclusions still hold if one includes the cross-susceptibilities; see the Supplementary Material of Ref.~\cite{Else_2010} for the precise arguments.)
Hence we find that
\begin{equation}
\delta \eta = \frac{m}{2\pi} \frac{\mathcal{E}}{\chi_{\Gamma\Gamma}},
\end{equation}

 Combining with \eqnref{eq:j_anomaly_1d}, we find that 
\begin{equation}
\label{eq:weight_1d}
\mathcal{W} = \frac{m^2}{4\pi}\frac{1}{\chi_{\Gamma\Gamma}}.
\end{equation}
The system is ``fluxible'' and ``compressible'' in the sense of section \ref{sec:compressibility} if the anomaly coefficient $m \neq 0$. Nevertheless, there is still a way to suppress the infinite conductivity, since it could be that the susceptibility $\chi_{\Gamma\Gamma}$ diverges. Indeed, one frequently expects divergent susceptibilities in systems exhibiting critical behavior. This therefore is the mechanism by which criticality can lead to nonzero DC resistivity in the IR fixed point theory, even in compressible systems.

One can derive similar formulas to \eqnref{eq:weight_1d} in the case of loop-group or 1-form symmetries in 2D. The loop group case was discussed extensively in Ref.~\cite{Else_2010} and we will not consider it further here. For 1-form symmetries we find (again assuming time-reversal symmetry, although similarly to before the main conclusions will still hold for the more general case) that the conductivity tensor takes the form
\begin{equation}
\sigma(\omega)^{ij} = \mathcal{W}^{ij} \delta(\omega) + \cdots,
\end{equation}
with
\begin{equation}
\mathcal{W}^{ij} = \frac{m^2}{4\pi} \epsilon^{ij} \epsilon^{kl} (\chi^{-1})_{jl},
\end{equation}
where $m$ is the anomaly coefficient, and  we defined the susceptibility tensor
\begin{equation}
\label{eq:sustensor}
\chi^{ij} = \frac{\partial}{\partial \eta_i} \langle \hat{u}^{j} \rangle_{\vec{\eta}} \biggr|_{\vec{\eta} = 0},
\end{equation}
where the expectation value is computed with respect to the generalized Gibbs ensemble \eqnref{eq:1_form_gibbs}, with $\vec{\eta}(\mathbf{x}) = \vec{\eta}$ independently of $\mathbf{x}$. Thus, to get suppression of the infinite DC conductivity with $m$ nonzero (the fluxible and compressible case), we would require that $\chi^{-1} = 0$, corresponding to divergence of the susceptibility matrix $\chi$.

\section{ An interesting solvable model: The Quantum Lifshitz critical point of bosons} 

\label{sec:qlm}

\subsection{Introducing the model}
The very general considerations above are nicely illustrated by a simple model that we now describe. 
A superfluid phase of bosons admits gapless Goldstone excitations which are conveniently described in terms of a real scalar field $\phi$ which describes the phase of the boson.  In the usual superfluid phase where the bosons condense at zero momentum the Euclidean phase action takes the familiar form: 
\be
\label{eq:sfaction}
S_{sf} = \int d\tau d^2x \frac{\kappa}{2} (\partial_\tau \phi)^2 + \frac{\rho_s}{2} (\vec \nabla \phi)^2 + .....
\ee
Here $\kappa$ is the bulk compressibility of the superfluid, and $\rho_s$ is the phase stiffness. The ellipses represent higher derivative and/or non-linear terms. 
The transition to a condensate at non-zero momentum occurs when the phase stiffness $\rho_s$ changes sign. The critical point itself corresponds to $\rho_s = 0$.  It is important then to keep higher derivative terms. The critical action thus reads 
\be
\label{QLM} 
S =  \int d\tau d^2x \frac{\kappa}{2} (\partial_\tau \phi)^2 + \frac{K}{2} (\nabla^2 \phi)^2 +  u (\vec \nabla \phi)^4 + .....
\ee
This is known as the Quantum Lifshitz Model (QLM) and has been studied in detail in the literature. A possible realization in cold atoms was proposed in Ref. \cite{po2015two}, and in twisted trilayer graphene in Ref. \cite{lake2021re}. The quadratic part of the action describes a scale invariant theory with dynamical critical exponent $z = 2$.  Interestingly this quadratic theory has a fixed line parameterized by $K$ throughout which the boson field $b \sim, e^{i\phi}$ has power law correlations. We have retained a non-linear term that is marginal by power-counting along the fixed line. It is marginally irrelevant for $u > 0$.

The QLM also appears as the critical theory describing phase transitions between  valence bond solid phases of $2d$ quantum magnets\cite{vishwanath2004quantum,fradkin2004bipartite}. In that context there is no microscopic global $\UU(1)$ symmetry corresponding to the conservation of the boson number conjugate to the phase.  Then terms that involve $\cos(n\phi)$ for some integer $n$
 are allowed in the action; these terms are (ir)relevant depending on the value of $K$. In contrast when the QLM arises at the critical point of a boson system with a microscopic global $\UU(1)$ symmetry, such $\cos(n\phi)$ terms are explicitly forbidden, and we have a stable fixed line for any value of $K$. 

The QLM transition in the superfluid can occur at any value of the microscopic density. As we will see shortly correspondingly the QLM theory has a finite compressibility.  Nevertheless, unlike a superfluid, it does not spontaneously break the global $\UU(1)$ symmetry of the microscopic boson system. 

The action in \eqnref{QLM} has continuum translation and rotational invariance.  It can be realized in a microscopic model with continuum translation and rotational symmetries. A useful example described in Ref. \cite{po2015two}  is a model of Rashba spin-orbit  coupled bosons. Alternately we can consider the same model with bosons sitting at the sites of a triangular lattice with  6-fold $C_6$ rotational symmetry.  This discrete $C_6$  symmetry is sufficient to allow only\footnote{In contrast on the square lattice other terms of the same order are allowed, and generically the transition is driven weakly first order. See Refs.~\cite{vishwanath2004quantum,fradkin2004bipartite}.} the terms explicitly included in the action in \eqnref{QLM}. Higher order derivative terms will break the continuous rotational symmetry to $C_6$ but they are irrelevant at low energies. 

In thinking about this model, we must recognize that -- as in the superfluid -- the effective theory of the QLM does not include any vortex excitations. This is because the effective theory describes the physics at energy scales below the vortex gap. However, if we put the system on a torus, there can still be non-trivial winding number of the phase field along the non-contractible cycles.
This amounts to saying that the 1-form symmetry $\mathrm{U}(1)_1$ that emerges in the superfluid (as previously discussed in Section \ref{subsec:1form}) is still present in the QLM. Moreover, this 1-form symmetry and the electric charge $\mathrm{U}(1)$ will have a mixed anomaly for the same reasons we previously described in the superfluid case.

As in a superfluid, the microscopic translation symmetry in the QLM is realized in a trivial way on the field $\phi$; it simply goes to itself. Nevertheless, in the superfluid and in the QLM, there is still a sense in which the microscopic translation symmetry is non-trivial with respect to objects in the theory. Specifically, although the theory does not intrinsically contain vortices, we can reintroduce them as extrinsic defects. As can be understood from invoking a charge-vortex duality for example, the vortices see the background microscopic electric charge density as an effective magnetic field. Therefore, the microscopic translation symmetry acts \emph{projectively} on these defects. Formally this corresponds to a non-trivial interplay between the microscopic translation symmetry and the emergent 1-form symmetry (this is an example of ``symmetry fractionalization''). One can moreover argue that due to the mixed anomaly between the electric charge $\mathrm{U}(1)$ symmetry and the 1-form symmetry, such a projective action of microscopic translation symmetry on the vortices implies that the microscopic filling $\nu$ is nonzero. (Specifically, the physical content of the argument is that the mixed anomaly amounts to the statement that a $2\pi$ magnetic flux binds a vortex, and therefore is acted upon projectively by the microscopic translation symmetry. The projective action of translations on a $2\pi$ magnetic flux determines the filling according to the general perspective of Ref.~\cite{Else_2007}). This is the analog of Luttinger's theorem in the context of superfluids and QLM.

The projective action of translations on the vortices, and hence the filling, can be continuously tuned, hence the superfluid and QLM are compressible states. They are also ``fluxible'' states in the sense of Section \ref{sec:compressibility}, for the reasons described in Section \ref{subsec:1form}, in line with the general conjecture of Section \ref{sec:compressibility} that fluxibility and compressibility are equivalent. For superfluids this lines up with the well-known fact that a superfluid has infinite conductivity. Below, we will describe the electrical transport properties of the QLM critical point, and we will find that the fixed point theory itself will have zero conductivity despite the fluxibility. This will be explained as a concrete and simple realization of the critical drag mechanism. However (dangerously) irrelevant perturbations will render the conductivity non-zero at non-zero temperatures.

\subsection{Physical properties} 
We have already mentioned that the QLM is compressible according to the definition used in this paper. An alternative definition of compressibility is simply that the thermodynamic compressibility $dn/d\mu$ is nonzero. We can easily verify that the QLM is also compressible according to this definition. To see this, we turn on  a chemical potential $\mu$ . This changes the time derivative term in the action to 
\be
\frac{\kappa}{2} (\partial_\tau \phi - i \mu)^2 
\ee
Clearly the susceptibility to $\mu$ is  simply given by $\kappa$ as promised. Thus the QLM critical point has $dn/d\mu \neq 0$.

Next consider the particle number conductivity. This may be discussed by turning on a transverse vector potential $\vec A_T$ that satisfies $\vec \nabla \cdot \vec A_T = 0$.  If $A_T$ has a wavenumber $\vec q$ and frequency $\omega$, the frequency-dependent conductivity $\sigma(\omega)$ requires considering the limit $\vec q \rightarrow 0$.  In the low energy QLM fixed point theory (where we set the irrelevant perturbation $u = 0$), the transverse gauge field does not couple to the action. Thus we conclude that, for the fixed point theory, the conductivity $\sigma(\omega) = 0$ for all $\omega$. Thus, despite being compressible, the QLM fixed point is an insulator at $T = 0$.

Let us now show how this result may be understood as an illustration of the critical drag mechanism. We argued in the previous section that the only way to suppress the dissipationless current is to have the susceptibility of the conserved charges of the $1$-form symmetry diverge. In the present example this conserved density takes the form 
\be
\label{1frmgen}
u^i = \frac{1}{2\pi} \epsilon^{ij} \partial_j \phi.
\ee
By completing the square in the action \eqnref{eq:sfaction}, one can compute the susceptibility tensor $\chi^{ij}$ defined in \eqnref{eq:sustensor}. The result is that
\begin{equation}
\chi^{ij} = \frac{1}{4\pi^2 \rho_s} \delta^{ij}.
\end{equation}
Clearly this susceptibility diverges at the QLM critical point where $\rho_s \to 0$, and this suppresses the dissipationless current. Thus we see that the QLM explicitly demonstrates the critical drag mechanism.

In the QLM we can also study the effect of irrelevant perturbations. We will see below that these will have important effects on the transport. Hence these perturbations should be regarded as dangerously irrelevant. 
First at zero temperature, the irrelevant perturbations will lead to a   non-zero finite frequency conductivity. In particular, $\vec A_T$ will couple to the theory through the $u$ term in \eqnref{QLM}, and this will lead to a nonzero $\sigma(\omega)$ that nevertheless will vanish in the dc limit $\omega \to 0$.

Let us now consider non-zero $T$. We begin by ignoring the vortices. Then at any non-zero $T$, we claim that the model has an infinite DC conductivity. This is because though the phase stiffness $\rho_s$ is tuned to zero at the $T = 0$ critical point, at non-zero $T$, the non-linear interaction(the $u$-term) will lead to a non-vanishing $\rho_s(T)$. This effect was calculated in Ref.~\cite{ghaemi2005finite}, and the result is 
\be
\label{rhosT} 
\rho_s(T) = c  \frac{T\sqrt{K}  \ln(\ln(\frac{1}{T}))}{\ln(\frac{1}{T})}
\ee
where $c$ is a number of order $1$. 
 The non-zero phase stiffness leads to a finite temperature complex conductivity 
\be
\sigma(\omega, T) = \frac{\rho_s(T)}{i\omega}
\ee
whose real part is a delta function in frequency. Hence the DC conductivity is infinity at any non-zero temperature.  In the general language of critical drag, this is because the 1-form susceptibility $\chi^{ij}$ becomes finite at nonzero temperature. In the absence of vortices, the system is a quasi-long range ordered superfluid at any $T > 0$. 

Now let us include the gapped vortices.  Note that as $\frac{\rho_s}{T}$  goes to zero as  $T$ goes to zero,  the quasi-long range ordered state is always unstable to proliferation of unbound vortex-antivortex pairs at  low $T$. Equivalently we are above the Berezinski-Kosterlitz-Thouless (BKT) transition temperature $T_{BKT}$ which, for a system with phase stiffness $\rho_s$ will occur at $T_{BKT} =  a\rho_s$ where $a$ is a constant of order $1$.   The deviation from this putative BKT transition may be parametrized by $t = \frac{T - T_{BKT}}{T_{BKT}} \approx \ln(\frac{1}{T})$ at low-$T$, which is hence not small. Consequently in what follows we will simply treat inter-vortex interactions as weak, and roughly estimate the free vortex density  $ n_v$   to be 
 \be
n_v \sim e^{-\frac{ \Delta}{k_B T} } 
\ee
where $\Delta$ is the vortex gap. (As a consequence of the nonzero vortex density at nonzero temperature, the 1-form symmetry is weakly broken. Therefore, it no longer perfectly protects the current and one expects nonzero resistivity regardless of the 1-form susceptibility). This will destroy the power law finite-$T$ long range order, and restore a finite DC conductivity. To estimate this, we follow the standard analysis\cite{halperin1979resistive} of electrical transport above the BKT transition in two dimensional superconductors.  The frequency dependent conductivity can then be expressed in the Drude form 
\be
\label{sigmafT}
\sigma(\omega, T) = \frac{D(T)}{\pi (-i \omega + \gamma (T) )} 
\ee
where the Drude weight $D(T) = \rho_s(T)$, and $\gamma(T)  \sim n_v (T) $.  Since $n_v$ goes to zero exponentially as $T \to 0$ while $\rho_s(T)$ only scales like a power law, we find that the DC conductivity scales (up to power law corrections) as
\be
\sigma(\omega = 0, T) \sim e^{\frac{\Delta}{k_B T}}
\ee
 We may understand this as the inverse of the vortex conductivity $\sigma_v = n_v \mu_v$ where $\mu_v$ is the mobility of the vortices. In the dc limit, we expect that the mobility will come from vortices scattering off the gapless phase fluctuations which will lead to some power law $T$-dependence of $\mu_v$. The temperature dependence will thus be dominated by that of $n_v$.
 
 We see therefore that there is a non-zero DC conductivity at non-zero $T$ which, furthermore, goes to infinity rapidly as $T \rightarrow 0$ despite the fact that the conductivity is \emph{zero} exactly at $T=0$. However, we should not interpret this fact as signifying that the critical drag mechanism is ineffective at nonzero temperature. Indeed, the \emph{weight} $D(T)$ of the Drude peak in \eqnref{sigmafT} does go to zero continuously as $T \to 0$, reflecting the critical drag.
The diverging DC conductivity arises because the width $\sim n_v$ of the Drude peak goes to zero even faster (exponentially). Thus to observe the large DC conductivity would require observing the system over exponentially long time scales in order to have enough frequency resolution to resolve the Drude peak. If the Drude peak is not resolved then one will instead measure the weight $D(T)$, which goes to zero as $T \to 0$. Another way to say this is that if one switches on a static electric field $E$ at time $t=0$, then the current $j(t)$ in the system will gradually increase with time at a rate $D(T) E$ until it finally saturates at a value of $\sigma_{\mathrm{DC}}(T) E$. At low temperatures where $D(T)$ is small, this increase will happen very slowly even if the final saturation value is large.
The behavior of the conductivity in this system is a dramatic illustration of a familiar phenomenon at quantum critical points where the $\omega \rightarrow 0$ and $T \rightarrow 0$ limits often do not commute. In the QLM model these two limits yield infinitely different answers!

It is also instructive to recast these results in the language of the emergent one-form symmetry and its weak breaking at finite temperature. We begin with the anomaly equation [\eqnref{sfanom}) for the emergent one-form symmetry which we have argued will remains present in the QLM fixed point. Specializing to spatially uniform situations, we focus on the time evolution of $u^i =  \mathcal{J}^{0 i}$ to write 
\be
\partial_0  u^i  = \frac{1}{2\pi} \epsilon^{ij} E_j 
\ee
At a small non-zero temperature thermally excited vortices will weakly relax the one-form density $u^i$ at the rate $\gamma \sim n_v$ . We describe this by adding a relaxation term to write 
\be
\partial_0 u^i + \gamma u^i = \frac{1}{2\pi} \epsilon^{ij} E_j 
\ee
For small deviations from equilibrium we write $u^i = \chi \eta^i$ where $\chi \sim \frac{1}{\rho_s(T)}$ is the 1-form susceptibility and $\eta^i$ is the one-form chemical potential.  Using now the result that the electrical current is $j^i \simeq \frac{m}{2\pi} \epsilon^{ij} \eta_j$, we immediately reproduce the conductivity of \eqnref{sigmafT}. 

What we can extrapolate from the above analysis  is that a conductivity of the form \eqnref{sigmafT}, with the Drude weight $D(T)$ and relaxation rate $\gamma(T)$ both going to zero as $T \to 0$, is likely generic in systems exhibiting critical drag, with the caveat that in general it only represents the ``coherent'' conductivity that is related to nearly conserved quantities, and in general there could be another additive contribution $\sigma_Q$ representing the ``incoherent'' conductivity \cite{Hartnoll_0706,Lucas_1502}. (In the QLM the incoherent contribution would only come from irrelevant terms and goes to zero in the DC limit, but the same need not be true in other systems.)
The Drude weight going to zero as $T \to 0$ results from the susceptibility of the appropriate (nearly) conserved quantity going to zero, while $\gamma(T)$ going to zero as $T \to 0$ is a consequence of the emergent symmetry. In the QLM example, $\gamma(T)$ goes to zero exponentially with $T$ and hence dominates the scaling of the coherent DC conductivity $\sigma_{\mathrm{DC}}^{\mathrm{coherent}}(T) = D(T)/\gamma(T)$; however, in the ersatz Fermi liquids discussed in Ref.~\cite{Else_2010}, it is more likely that $D(T)$ and $\gamma(T)$ both scale as some power law in $T$. The corresponding exponents will determine how $\sigma_{\mathrm{DC}}^{\mathrm{coherent}}(T)$ scales as $T \to 0$; it does not \emph{necessarily} diverge as it does in the QLM. In any case, as previously mentioned in Ref.~\cite{Else_2010}, in ersatz Fermi liquids, weak disorder can also likely lead to $\gamma(T)$ going to a nonzero value as $T \to 0$. Combined with $D(T)$ going to zero, this would certainly ensure that $\sigma_{\mathrm{DC}}^{\mathrm{coherent}}(T)$ goes to zero as $T \to 0$.

\section{Discussion} 
\label{sec:discussion}
In this work we have presented a general point of view on the origin of nonzero resistivity and its relation with conserved quantities. In particular, we have argued that in compressible systems, there is always an emergent conserved quantity that would protect the current and lead to infinite DC conductivity in the RG fixed point theory, \emph{unless} the weight of the Drude peak is suppressed by critical fluctuations, i.e.\ the mechanism we have called ``critical drag''. Moreover, we have presented a solvable model of critical drag, namely the quantum Lifshitz model (QLM).

The QLM itself may be relevant to certain experiments such as the re-entrant superconductivity in twisted trilayer graphene \cite{lake2021re}. However, in this paper we mainly intended to use the QLM as proof of principle for the critical drag mechanism. On the other hand, for the reasons described here and previously in Ref.~\cite{Else_2010}, it seems likely that the ``strange metal'' seen in cuprates and heavy fermion materials must also manifest critical drag. It is therefore a very important open question to develop models that exhibit critical drag while being more relevant to such systems.

In this work we have focused on the electrical conductivity. However, critical drag may also have implications for thermal and thermoelectric transport coefficients, as well as the electrical transport in the presence of magnetic fields. We leave such developments for future work.


\begin{acknowledgments}
D.V.E.\ was supported by the Gordon and Betty Moore Foundation, grant nos.\ GBMF8683 and GBMF8684. T.S.\ is supported by a US Department of Energy grant DE- SC0008739, and in part by a Simons Investigator award from the Simons Foundation. This work was also partly supported by
the Simons Collaboration on Ultra-Quantum Matter, which is a grant from the Simons Foundation (651440, TS).

\end{acknowledgments}

\appendix

\section{Generalized Gibbs ensemble for higher-form symmetries}
\label{appendix:generalized_gibbs}

Here we will give the derivation of the thermal equilibrium state in the presence of higher-form symmetries. For pedagogical reasons, we first present the argument for the particular case of a 1-form symmetry for a system on the 2-dimensional torus, and then we will generalize to a $k$-form symmetry on an arbitrary oriented manifold, at the cost of introducing additional abstraction.

\subsubsection{1-form symmetry on the 2-torus}
Consider a system with a 1-form symmetry for which space is a 2-dimensional torus with extent $L_x$ and $L_y$ in the $x$ and $y$ directions respectively. Then as described in Section \ref{subsec:1form}, there is a ``charge density'' $\hat{u}^i$, which is a contravariant spatial vector that satisfies
\begin{equation}
\label{eq:u_divergence_appendix}
\partial_i \hat{u}^i = 0.
\end{equation}

The non-trivial conserved charges are found by integrating $\hat{u}^i$ over non-trivial cycles of the torus, i.e.
\begin{equation}
\label{eq:Qy}
\hat{\mathcal{Q}}^y = \int_0^{L_x} \hat{u}^y(x,a) dx,
\end{equation}
and
\begin{equation}
\hat{\mathcal{Q}}^x = \int_0^{L_y} \hat{u}^x(a,y) dy,
\end{equation}
where \eqnref{eq:u_divergence_appendix} ensures that the results of the integrals are independent of $a$.

In particular, if we replace $a$ with $y$ in \eqnref{eq:Qy} and average over $y$, we find another expression for $\hat{\mathcal{Q}}^y$ (and similarly for $\hat{\mathcal{Q}}^x$). We conclude that
\begin{equation}
\hat{\mathcal{Q}}^i = \frac{1}{L_i} \int \,  \hat{u}^i(\mathbf{x}) d^2 \mathbf{x}
\end{equation}
[no implicit summation] for $i = x,y$, where the integral is over the whole torus.

Now we want to find the state that maximizes the von Neumann entropy subject to the constraint that the expectation values of $\hat{\mathcal{Q}}^x$ and $\hat{\mathcal{Q}}^y$, as well as the total energy and electric charge $\hat{Q}$, are held fixed. By standard arguments, such a state is given by
\begin{equation}
\label{eq:rho_torus}
\rho = \frac{1}{\mathcal{Z}} \exp\left(-\beta \left[ H - \mu \hat{Q} - \lambda_i \hat{\mathcal{Q}}^i \right] \right)
\end{equation}
for some Lagrange multipliers $\lambda_x, \lambda_y$. If we now define the spatial vector $\eta_i = \lambda_i/L_i$ (no implicit summation), then \eqnref{eq:rho_torus} becomes
\begin{equation}
\rho = \frac{1}{\mathcal{Z}} \exp\left(-\beta \left[ H - \mu \hat{Q} - \int \eta_i u^i(\mathbf{x}) d^2 \mathbf{x} \right] \right),
\end{equation}
in agreement with \eqnref{eq:1_form_gibbs}. In this case $\eta$ can be taken to be a \emph{constant} 1-form, which certainly implies that it is closed.

\subsubsection{The general case}

Consider a system with a $k$-form symmetry in $d$ spatial dimensions. Then there is a ``charge density'' $\hat{u}^{i_1 \cdots i_k}$, which is an a rank-$k$ antisymmetric tensor that satisfies 
\begin{equation}
\label{eq:divergence_of_charge_density}
\partial_{i_1} \hat{u}^{i_1 \cdots i_k} = 0.
\end{equation}
From this we can define a dual $(d-k)$-form $\hat{\xi}$ with components 
\begin{equation}
\label{eq:xi_rho}
\hat{\xi}_{j_1 \cdots j_{d-k}} = \epsilon_{j_1 \cdots j_{d-k} i_1 \cdots i_k} \hat{u}^{i_1 \cdots i_k}.
\end{equation}
\eqnref{eq:divergence_of_charge_density} implies that $\hat{\xi}$ is a closed $(d-k)$-form, $d\hat{\xi} = 0$.
The charge $\hat{Q}_\Sigma$ associated with a closed $(d-k)$-dimensional surface $\Sigma$ is given by
\begin{equation}
\hat{Q}_\Sigma = \int_\Sigma \hat{\xi}.
\end{equation}
The fact that $\hat{\xi}$ is closed implies that $\hat{Q}_\Sigma$ only depends on the homology class of $\Sigma$. In fact, $\hat{Q}_\Sigma$ is still conserved when $\Sigma$ is any $(d-k)$-chain with coefficients in $\mathbb{R}$.

Next we invoke the standard fact, which follows from the theory of Poincar\'e duality \cite{Hatcher} (for  a concise and useful review, see  the Appendix of Ref.~\cite{lake2018higher}), that if the spatial manifold $X$ is oriented, then for every closed $(d-k)$-chain $\Sigma$, there is a corresponding closed $k$-form $\sigma_\Sigma$ such that
\begin{equation}
\label{eq:poincare}
\int_\Sigma \xi = \int_X \sigma_\Sigma \wedge \xi
\end{equation}
for any closed $(d-k)$-form on $X$. The map $\Sigma \mapsto \sigma_\Sigma$ induces an isomorphism from $H_{d-k}(X,\mathbb{R}) \to H^{k}(X, \mathbb{R})$.

Now we want to find the state that maximizes the von Neumann entropy subject to the constraint that the expectation values of $\hat{Q}_\Sigma$ for any closed $(d-k)$-chain $\Sigma$, as well as the total energy and electric charge, are held fixed. By \eqnref{eq:poincare}, it is equivalent to require that the expectation value of $\int_X \sigma \wedge \hat{\xi}$ is held fixed for every closed $(d-k)$-form $\sigma$. In turn this is equivalent to requiring that the expectation value of $\int_X \sigma_j \wedge \hat{\xi}$ is held fixed for each $j$, where $\sigma_1, \cdots \sigma_n$ are representative $k$-forms corresponding to a set of generators for $H^{k}(X, \mathbb{R})$. By standard arguments, the state that maximizes the entropy subject to these constraints is
\begin{equation}
\label{eq:lagrange}
\rho = \frac{1}{\mathcal{Z}} \exp\left(-\beta\left[\hat{H} - \mu \hat{Q} - \sum_{j=1}^n \lambda_j \int_X \sigma_j \wedge \hat{\xi}\right]\right).
\end{equation}
where $\lambda_1,\cdots,\lambda_n$ are real-valued Langrange multipliers. Now if we define the closed $k$-form $\eta = \sum_{j=1}^n \lambda_j \sigma_j$, then \eqnref{eq:lagrange} becomes
\begin{equation}
\rho = \frac{1}{\mathcal{Z}} \exp\left(-\beta\left[\hat{H} - \mu \hat{Q} - \int_X \eta \wedge \hat{\xi} \right]\right),
\end{equation}
If we invoke the relation \eqnref{eq:xi_rho}, then we can also write this as
\begin{equation}
\rho = \frac{1}{\mathcal{Z}} \exp\left(-\beta\left[\hat{H} - \mu \hat{Q} - \int_X \eta_{i_1 \cdots i_k} \hat{u}^{i_1 \cdots i_k}  d^d \mathbf{x}  \right] \right).
\end{equation}

\section{Proof of Bloch's theorem in d dimensions}
\label{appendix:bloch_general_dimension}
Consider a system that lives on a $d$-dimensional manifold $X$. For the simple case where $X$ is a torus, the argument for Bloch's theorem (without mixed anomalies) was discussed in Ref.~\cite{Watanabe_1904} and can easily be generalized to include mixed anomalies. Here we will give a more general argument for an arbitrary oriented manifold $X$. The result we need is that for any closed $(d-1)$-dimensional surface $\Sigma$ in $X$ [or generally any closed $(d-1)$-chain on $X$ with $\mathbb{Z}$ coefficients
, there exists a scalar function $\lambda_\Sigma$ on $X$ valued in $\mathbb{R}/(2\pi\mathbb{Z})$ [that is, $\lambda_\Sigma(\mathbf{x})$ is an angular variable] with the property that any divergenceless vector field $\mathbf{j}(\mathbf{x})$ satisfies
\begin{equation}
\label{eq:j_equivalence}
\int_\Sigma \mathbf{j} \cdot d\mathbf{\Sigma} = \frac{1}{2\pi} \int_X (\nabla \lambda) \cdot \mathbf{j}
\end{equation}
To show this one first invokes the Poincar\'e duality isomorphism previously described in Appendix A, which gives a closed 1-form $\sigma_\Sigma$ such that
\begin{equation}
\int_\Sigma \mathbf{j} \cdot d\mathbf{\Sigma} = \int_X (\sigma_\Sigma)_i j^i
\end{equation}
for any divergenceless vector field $\mathbf{j}$. Moreover, because $\sigma_\Sigma$ originated from an \emph{integer}-valued $(d-1)$-chain $\Sigma$, it follows that the integral of $\sigma_\Sigma$ over any closed cycle on $X$ is an integer. It follows that if we try to define an angular field $\lambda$ on $X$ through the equation
\begin{equation}
\frac{1}{2\pi} \partial_i \lambda = (\sigma_\Sigma)_i,
\end{equation}
then there exists a global solution. \eqnref{eq:j_equivalence} then follows.

Now if we define
\begin{equation}
U_{\lambda} = \exp\left(-i\int \lambda(\mathbf{x}) \hat{n}(\mathbf{x}) d^d \mathbf{x} \right),
\end{equation}
then we have analogously to \eqnref{eq:UHU} that
\begin{align}
U_{\lambda_\Sigma} \hat{H} U_{\lambda_\Sigma}^{-1} &= \hat{H} + 
\int [\nabla \lambda_\Sigma(\mathbf{x})] \cdot \hat{\mathbf{j}}(\mathbf{x}) + \cdots \label{eq:dots_2d} \\
&= \hat{H} + \hat{I}_\Sigma + \cdots,
\end{align}
where $\hat{I}_\Sigma = \int_\Sigma \hat{\mathbf{j}} \cdot d\boldsymbol{\Sigma}$ is the electric current flowing through $\Sigma$, and we have invoked \eqnref{eq:j_equivalence}. The arguments then proceed similarly to the 1D case. The main difference is in the estimating the contribution of the ``$\cdots$'' terms in \eqnref{eq:dots_2d}. We can roughly estimate their maximum possible contribution to the current as $O(\Gamma_\Sigma)$, where we have defined the geometrical factor $\Gamma_\Sigma$ by
\begin{equation}
\Gamma_\Sigma = \int R_\lambda(\mathbf{x})^{-2} d^d \mathbf{x},
\end{equation}
where $R_\lambda(\mathbf{x})$ is the radius of variation of $\lambda$ near the point $\mathbf{x}$. One can argue that $\Gamma_\Sigma$ is roughly bounded above by $|\Sigma|R^{-1}$, where $R$ is the length of the smallest loop in $X$ that has non-trivial intersection number with $\Sigma$ .

\section{Commutator of densities from 't Hooft anomaly in 1D}
\label{appendix:commutator}
We imagine subjecting the system to a time-dependent Hamiltonian $
\hat{H} + \hat{V}(t)$, where
\begin{equation}
\hat{V}(t) = \int \phi(x,t) \hat{n}(x),
\end{equation}
and $\phi(x,t)$ represents the scalar potential, such that
\begin{equation}
 -\partial_x \phi(x,t) = E(x,t)
\end{equation} is the electric field.

It follows that
\begin{equation}
\label{eq:Hcommutator_a}
\partial_t \hat{n}_\Gamma(x,t) \biggr|_{t=0} = i[\hat{H}, \hat{n}_\Gamma(x)] +i \int \phi(x',0) [\hat{n}(x'), \hat{n}_\Gamma(x)] dx'
\end{equation}
Now, from the continuity equation in the absence of applied electric field, we have that
\begin{equation}
\label{eq:Hcommutator_b}
i[\hat{H}, \hat{n}_\Gamma(x)] = -\partial_x \hat{j}_\Gamma(x),
\end{equation}
where $\hat{j}_\Gamma(x)$ is the current operator. Now if we combine \eqnref{eq:Hcommutator_a}, \eqnref{eq:Hcommutator_b}, and the anomaly equation \eqnref{eq:0form_anomaly}, we find that
\begin{equation}
\int \phi(x',0) [\hat{n}(x'), \hat{n}_\Gamma(x)] = i\frac{m}{2\pi} \partial_x \phi(x,0)
\end{equation}
Since this must hold for any functional form of $\phi(x,0)$, we conclude that
\begin{equation}
\label{eq:rho_rho_M_copy}
[\hat{n}(x), \hat{n}_\Gamma(x')] = -i \frac{m}{2\pi} \delta'(x - x'),
\end{equation}
which is \eqnref{eq:rho_rho_M}.

\section{Loop group anomalies}
\label{appendix:loop_group_anomalies}
The 't Hooft anomalies of the loop group $\mathrm{LU}(1)$ were discussed in Ref.~\cite{Else_2007}. Let us review the results.

 Firstly, a gauge field for $\mathrm{LU}(1)$ comprises a covariant vector $A_\mu$ on $M \times S^1$, where $M$ is the space-time manifold, and $S^1$ parameterizes the Fermi surface. The component of $A_\theta$ of $A$ along the Fermi surface can be interpreted as a Berry's phase for Fermi surface excitations. The current $J^\mu$ associated to the loop group conservation law is a contravariant vector field on $M \times S^1$, and contains a component $J^\theta$ describing the flow of charge along the Fermi surface.
 
The anomaly equation then takes the form
\begin{equation}
\label{eq:loop_group_anomaly}
\partial_\mu J^\mu = \frac{m}{8\pi^2} \epsilon^{\mu \nu \lambda \sigma} (\partial_\mu A_\nu)(\partial_\lambda A_\sigma),
\end{equation}
where the anomaly coefficient $m$ is quantized to be an integer.
From this one can derive, analogously to the 1-D case (see Section \ref{sec:emergent_1d} and Appendix \ref{appendix:commutator}) the commutation relation for the densities (see also Refs.~\cite{Golkar_1602,Barci_1805,Else_2007}):
\begin{equation}
[\hat{n}(\theta, \mathbf{x}), \hat{n}(\theta',\mathbf{x}')] = -i \frac{m}{(2\pi)^2} \epsilon^{abc} \partial_a A_b \partial_c [ \delta(\theta - \theta') \delta^2(\mathbf{x} - \mathbf{x}')],
\end{equation}
where the indices $a,b,c$ range over the two spatial dimensions and the $\theta$ dimension (but not time). If we assume that the magnetic field $\epsilon^{ij} \partial_i A_j$ is zero (where indices $i$ and $j$ range only over the spatial dimensions), then
 this reduces to
\begin{equation}
[\hat{n}(\theta, \mathbf{x}), \hat{n}(\theta',\mathbf{x}')] = -i \frac{m}{(2\pi)^2} \epsilon^{ij} F_{i\theta}(\theta) \delta(\theta - \theta') \partial_j \delta^2(\mathbf{x} - \mathbf{x}'),
\end{equation}
where we defined
\begin{equation}
F_{i\theta} = \partial_i A_\theta - \partial_\theta A_i.
\end{equation}
If we now repeat the argument for Bloch's theorem, we find that in the thermodynamic limit the electric current is given by
\begin{equation}
j^i = \frac{m}{(2\pi)^2} \int \epsilon^{ij} F_{j\theta}(\theta) \mu(\theta) d\theta.
\end{equation}

It remains to determine how $F_{i\theta}$ is related to microscopic quantities. Without a magnetic field, the right-hand side of the anomaly equation \eqnref{eq:loop_group_anomaly} reduces to $(m/4\pi^2) \epsilon^{ij} E_i F_{j\theta}$. On the other hand, in a Fermi liquid one can compute the charge nonconservation exactly, and one finds that the right-hand side of \eqnref{eq:loop_group_anomaly} should take the form $(1/4\pi^2) \epsilon^{ij} E_i \partial_\theta k_i(\theta)$. Hence, at least in a Fermi liquid (for which we already know that $m=1$), we must set
\begin{equation}
F_{i\theta} = \partial_\theta k_i(\theta).
\end{equation}
Presumably this relation still holds in any ersatz Fermi liquid; see for example Section VI.B of Ref.~\cite{Else_2007} for an argument based on dimensional reduction

\section{Proof of Conjecture in $d=1$.}
\label{appendix:compressibility}
Here we give the proof of the Conjecture of Section \ref{sec:compressibility} for 1-D systems. It is not possible to have non-trivial higher form symmetries in 1-D, so we can assume the emergent symmetry group is some 0-form symmetry group $G$; specifically we will take to be a finite-dimensional Lie group (not necessarily compact), with Lie algebra $\mathfrak{g}$.

To formulate the result, we first recall from the general filling framework of Ref.~\cite{Else_2007} that a 't Hooft anomaly of $G$ defines a function $\alpha(\mathcal{Q}|\tau)$ valued in $\mathbb{R}/\mathbb{Z}$ that is defined when $\mathcal{Q} \in \mathfrak{g}$ and $\tau \in G$ commute, in which case $\alpha(\mathcal{Q}|\tau)$ computes the microscopic electric charge filling (modulo 1) of a system with emergent symmetry $G$ in which the microscopic electric charge maps into $\mathcal{Q}$ and the microscopic translation symmetry maps into $\tau$. Physically, $\alpha(\mathcal{Q}|\tau)$ describes the charge of $\tau$ created by a $2\pi$ flux insertion of $\mathcal{Q}$. Here the only properties of $\alpha$ that we need are:
\begin{itemize}
\item $\alpha(\mathcal{Q}|\tau)$ is linear in $\tau$: that is, if $\tau$ and $\tau'$ commute with $\mathcal{Q}$, then
\begin{equation}
\alpha(\mathcal{Q}|\tau \tau') = \alpha(\mathcal{Q}|\tau) + \alpha(\mathcal{Q}|\tau'),
\end{equation}
\item $\alpha(\mathcal{Q}|\tau)$ is continuous in $\tau$,
and
\item $\alpha(\mathcal{Q}|\tau)$ is invariant under conjugation: that is, for any $g \in G$ we have that
\begin{equation}
\alpha(g \mathcal{Q} g^{-1} | g \tau g^{-1}) = \alpha(\mathcal{Q}|\tau).
\end{equation}
\end{itemize}

Now, consider $\mathcal{A},\mathcal{Q} \in \mathfrak{g}$ that commute. Then we say that $\mathcal{A}$ and $\mathcal{Q}$ have a \emph{mixed anomaly} if
\begin{equation}
\label{eq:computation_of_m}
m := \frac{d}{d\theta} \alpha(\mathcal{Q} | e^{i\theta \mathcal{A}}) \biggr|_{\theta = 0} \neq 0.
\end{equation}
This in turn implies that the local densities corresponding to $\mathcal{Q}$ and $\mathcal{A}$ have a non-trivial commutator of the form \eqnref{eq:rho_rho_M} with the coefficient $m$ given by \eqnref{eq:computation_of_m}.

The precise formulation of the Conjecture of Section \ref{sec:compressibility} for the case of one spatial dimension is that a system is compressible if and only if there is $\mathcal{A} \in \mathfrak{g}$ that commutes with the electric charge $\mathcal{Q}$ and has mixed anomaly with it.
One direction of the ``if and only if'' is easy to prove. Specifically, if $\mathcal{Q}$ and $\mathcal{A}$ commute and have a mixed anomaly, then for any $\tau \in G$, we can define a one-parameter deformation $\tau_\theta = \tau e^{i\theta \mathcal{A}}$, and we see that 
\begin{equation}
\alpha(\mathcal{Q}|\tau_\theta) =  \alpha(\mathcal{Q}|\tau) + m\theta.
\end{equation}
Hence, the filling of the system can be continuously tuned by varying how the microscopic translation embeds into $G$, so the system is compressible.

The converse is a little more challenging and constitutes

\begin{thm}
\label{thm:compressibility}
Suppose $G$ is a finite-dimensional Lie group, with Lie algebra $\mathfrak{g}$, and a fix a function $\alpha$ satisfying the properties described above.
Let $S \subseteq \mathfrak{g}$ comprise those $\mathcal{Q}$ such that $e^{2 \pi i \mathcal{Q}} = 1$. Then
there exists a set $C \subseteq \mathbb{R}/\mathbb{Z}$ such that $(\mathbb{R}/\mathbb{Z}) \setminus C$ is countable, and for any $\mathcal{Q} \in S$ and $\tau \in G$ such that $\mathcal{Q}$ and $\tau$ commute and satisfy $\alpha(\mathcal{Q} | \tau) \in C$, it follows that there exists $\mathcal{A} \in \mathfrak{g}$ that commutes with $\mathcal{Q}$ and has mixed anomaly with $\mathcal{Q}$.
\end{thm}

Note in particular that $(\mathbb{R}/\mathbb{Z}) \setminus C$ countable implies that any open interval $(\nu_1, \nu_2)$ contains uncountably infinite number of points in $C$. Therefore, if the system is compressible, then we can tune the filling to lie in $C$ and we conclude that there is an $\mathcal{A} \in \mathfrak{g}$ that has mixed anomaly with $\mathcal{Q}$.

To prove Theorem \ref{thm:compressibility} we will make use of the following Lemma, whose proof we give later:
\begin{lemma}
\label{lem:countable_set}
 There exists a countable set $S_* \subseteq S$ such that for any $\mathcal{Q} \in S$, there exists $u \in G$ such that $u \mathcal{Q} u^{-1} \in S_*$.
\end{lemma}

Now, for any $\mathcal{Q} \in S$, let us define $H_{\mathcal{Q}} \leq G$ to comprise those elements that commute with $\mathcal{Q}$. $H_{\mathcal{Q}}$ is a Lie subgroup of $G$. Therefore, like any Lie group, or indeed any manifold, it has countably many connected components. This follows from the second-countability property that is part of the definition of a manifold.

Now define
\begin{equation}
V_{\mathcal{Q}}^{(0)} := \{ \alpha(\mathcal{Q}|\tau) : \tau \in H_{\mathcal{Q}}^{(0)} \}
\end{equation}
where $H_{\mathcal{Q}}^{(0)}$ is the connected component of the identity in $H_{\mathcal{Q}}$.
Observe that $V_{\mathcal{Q}}^{(0)}$ is the image under a continuous map of a connected set; it follows that $V_{\mathcal{Q}}^{(0)}$ is itself connected. The only connected subsets of $\mathbb{R}/\mathbb{Z}$ are intervals. If $V_{\mathcal{Q}}^{(0)}$ is an interval $I \subseteq \mathbb{R}/\mathbb{Z}$ of nonzero length, then since on the one other hand we see from the definition $n V_{\mathcal{Q}}^{(0)} \subseteq V_{\mathcal{Q}}^{(0)}$, while on the other hand $nI = \mathbb{R}/\mathbb{Z}$ for sufficiently large $n$; hence, we find that $V_{\mathcal{Q}}^{(0)} = \mathbb{R}/\mathbb{Z}$. The other possibility is that $V_{\mathcal{Q}}^{(0)}$ is a single point.

Let $\Sigma \subseteq S$ comprise those $\mathcal{Q}$ such that $V_{\mathcal{Q}}^{(0)}$ is a single point, and define $\Sigma_* = \Sigma \cap S_*$, and
\begin{equation}
\mathbb{R}/\mathbb{Z} \supseteq C^c := \{ \alpha(\mathcal{Q} | \tau) : \mathcal{Q} \in  \Sigma , \tau \in H_{\mathcal{Q}} \}.
\end{equation}
Then from Lemma \ref{lem:countable_set} we see that we can equivalently write
\begin{equation}
C^c := \{ \alpha(\mathcal{Q} | \tau) : \mathcal{Q} \in  \Sigma_* , \tau \in H_{\mathcal{Q}} \}.
\end{equation}
But now for any $\mathcal{Q} \in \Sigma_*$,  $\alpha(\mathcal{Q}|\tau)$ only depends on the connected component of $\tau$ in $H_{\mathcal{Q}}$. Since $\Sigma_*$ is countable and $H_{\mathcal{Q}}$ only has countably many connected components (see above), it follows that $C^c$ is countable.

Now, if we have some $\mathcal{Q} \in S$, $\tau \in G$ commuting such that $\alpha(\mathcal{Q}|\tau) \in C := (\mathbb{R}/\mathbb{Z}) \setminus C^c$, it follows that $\mathcal{Q} \notin \Sigma$. Therefore, $V_{\mathcal{Q}}^{(0)} = \mathbb{R}/\mathbb{Z}$. So if we define a Lie group homomorphism $\varphi : H_{\mathcal{Q}}^{(0)} \to \mathbb{R}/\mathbb{Z}$ according to $\varphi(\tau') = \alpha(\mathcal{Q}|\tau')$, then $\varphi$ is surjective. It follows that the Lie algebra homomorphism $\widetilde{\varphi} : \mathrm{Lie}(H_{\mathcal{Q}}) \to \mathbb{R}$ induced by $\varphi$ (where $\mathrm{Lie}(H_{\mathcal{Q}}) \leq \mathfrak{g}$ is the Lie algebra of $H_{\mathrm{Q}}$)
must be surjective, and in particular not zero; hence, there exists some $\mathcal{A} \in \mathrm{Lie}(H_{\mathcal{Q}})$ such that $\widetilde{\varphi}(\mathcal{A}) \neq 0$, which implies that $\mathcal{A}$ commutes with $\mathcal{Q}$ and has mixed anomaly with $\mathcal{Q}$. This completes the proof of Theorem \ref{thm:compressibility}.

\subsubsection{Proof of Lemma \ref{lem:countable_set}}
Here we give the proof of Lemma \ref{lem:countable_set}. This follows as a consequence of the following standard theorems \cite{Stroppel,Hall}:

\begin{thm}
\label{thm:compact_subgroup}
Let $G_0$ be a locally compact connected group. Then every compact subgroup of $G_0$ is contained in a maximal compact subgroup of $G_0$, and all maximal compact subgroups of $G_0$ are in the same conjugacy class.
\end{thm}

\begin{thm}
\label{thm:torus}
Let $G$ be a compact finite-dimensional Lie group. Then every torus in $G$ is contained in a maximal torus, and all maximal tori of $G$ are in the same conjugacy class.
\end{thm}
(A ``torus'' in a Lie group $G$ is a Lie subgroup that is isomorphic to $\mathrm{U}(1)^{\times k}$ for some $k$.)

By combining these two results we obtain
\begin{cor}
\label{cor:thecor}
Let $G$ be a finite-dimensional Lie group. Then every torus $T \leq G$ is contained in a torus $\mathbb{T}[T] \leq G$, such that $\mathbb{T}[T]$ and $\mathbb{T}[T']$ are in the same conjugacy class for any tori $T,T' \leq G$.
\begin{proof}
Let $T$ be some torus of $G$. Then clearly $T \leq G_0$, where $G_0$ is the connected component of the identity in $G$. Hence, applying Theorem \ref{thm:compact_subgroup} (recalling that finite-dimensional Lie groups are always locally compact), we find that $T$ is contained within a maximal compact subgroup $\mathbb{H}[T] \leq G_0$. The closed subgroup theorem implies that $\mathbb{H}[T]$ is itself a finite-dimensional compact Lie group. Hence, applying Theorem \ref{thm:torus} with respect to $H$, we find that $\mathbb{H}[T]$ is contained in a maximal torus of $H$, which we call $\mathbb{T}[T]$. If we start from two tori $T,T' \leq G$, then Theorem \ref{thm:compact_subgroup} shows that there exists $g \in G$ such that $g \mathbb{H}[T] g^{-1} = \mathbb{H}[T']$. Then since $g \mathbb{T}[T] g^{-1}$ and $\mathbb{T}[T']$ are both maximal tori in $\mathbb{H}[T']$, Theorem \ref{thm:torus} shows that they are in the same conjugacy class.
\end{proof}
\end{cor}

Now let $T_*$ be a representative of the unique conjugacy class of tori $\mathbb{T}[T]$ constructed in Corollary \ref{cor:thecor}. Let $\mathfrak{t} \leq \mathfrak{g}$ be its Lie algebra. Then the intersection $S_* := \mathfrak{t} \cap S$ is countable, since any $\mathcal{Q} \in S_*$ can be expressed as
\begin{equation}
\mathcal{Q} = \sum_{j=1}^n k_j \mathcal{Q}_j,
\end{equation}
for some integers $k_j$, and where $\mathcal{Q}_1 \cdots \mathcal{Q}_n$ are the standard generators of the torus $T_*$. Then Lemma \ref{lem:countable_set} follows.

\section{Bose-Luttinger Liquids: emergent symmetries, anomalies, and transport} 
\label{appendix:BLL} 
In this Appendix, we briefly discuss the Bose-Luttinger Liquids (BLL) introduced in Ref.~\cite{lake2021bose} and show how they fit in with the considerations of the present paper.  For simplicity we restrict to space dimension $d = 2$.  The BLL is a phase of matter of bosons, possibly at finite density, which  has gapless excitations living on a surface in momentum space.  We will consider such a phase in a microscopic system of bosons with global $\UU(1)$ and continuous translation symmetries.  The BLL ground state preserves these symmetries. Further it can exist at non-zero boson density, and is a compressible phase. The microscopic boson field $\psi_{\bf x}$ is expressed in terms of `patch' fields $\psi_\gamma$ that live near the Bose surface (a circle of radius $k_B$): 
\be \label{patch_decomp} \psi({\bf x}) = \frac{1}{\sqrt N}\sum_\gamma e^{ik_B \bf \gamma \cdot \bf x} \psi_\gamma(\bf x),\ee 
Here $N$, the number of patches, is to be taken to infinity (and correspondingly the size of each patch to zero). 

In the IR theory these patch fields are described in terms of their phase fields 
\be
\psi_\gamma \sim e^{i\phi_\gamma} 
\ee
The action of the IR theory is quadratic in terms of $\phi_\gamma$. The resulting IR fixed point has power law correlations of the boson $\psi_{\bf x}$ everywhere on the Bose surface. 

As noted in Ref.~\cite{lake2021bose} the BLL has a huge emergent symmetry group. First there is a loop group $\LU(1)$ symmetry associated with conservation of boson number at each point of the Bose surface. Our interest here is in emergent higher form symmetries associated with vortex excitations of the boson field which were argued\cite{lake2021bose} to be gapped at the IR fixed point.  This prompted the suggestion in Ref.~\cite{lake2021bose} that there was an emergent one-form symmetry (similar to the superfluid or the QLM examples).  Here we shall argue that there is in fact an emergent two-form symmetry group (which we denote $(\UU(1))_2$). To see this we first observe that we can view the $\gamma$ direction as an additional spatial dimension (see Ref.~\cite{Else_2007} for an analogous viewpoint on  Landau Fermi liquids). In other words we will regard the patch fields $\phi(\bf x, \gamma, \tau)$  as living in a $3+1$ dimensional space-time with $\bf x, \gamma$ being the `spatial' coordinates and $\tau$ the time coordinate. We define the conserved currents 
\be
{\mathcal J}^{\mu \nu \lambda} = \frac{1}{2\pi} \epsilon^{\mu\nu\lambda\kappa}\partial_\kappa \phi
\ee
where the Greek indices $\mu, \nu, \lambda, \kappa$ all take values in $\tau, x, y, \gamma$. 

These conserved currents define a two-form  $(U(1))_2$ symmetry.  There are three corresponding conserved densities: 
\begin{eqnarray} 
J^{0\gamma x} & = & \frac{1}{2\pi} \partial_y \phi(\bf x, \gamma, \tau) \\
J^{0y\gamma} & = & \frac{1}{2\pi} \partial_x \phi( \bf x, \gamma, \tau) \\
J^{0xy} & = &  \frac{1}{2\pi} \partial_\gamma \phi( \bf x, \gamma, \tau) 
\end{eqnarray} 
To understand the meaning of these conserved densities, we place the system on a spatial torus. (Note that the $\gamma$ direction naturally lives in $S^1$; so we are only requiring that $x$ and $y$ also be taken to each live in $S^1$.) Then we may define 3 distinct integer winding numbers 
\be
W_i = \int dx_i \frac{1}{2\pi} \partial_i \phi(\bf x, \gamma, \tau)
\ee
for $i = x, y, \gamma$. Consider first $W_x$ or $W_y$ . This represents the winding number of $\phi({\bf x}, \gamma, \tau)$ at any particular $y, \gamma$. As in previous sections, this winding number is independent of the $y$-coordinate. Crucially it is also independent of $\gamma$. This is because of, as noted in Ref.~\cite{lake2021bose},  the requirement of smoothness of the $\phi_\gamma$ on the Bose surface. Indeed we can simply regard $W_x$ as the winding number (along the x-direction) of the phase of the full boson field $\psi(\bf x)$.  This leads to a winding of the phase of all of the patch bosons.  Next consider $W_\gamma$. This represents the winding of the phase on moving around the Bose surface. Configurations with non-zero integer $W_\gamma$ are well-defined and are smooth on the Bose surface, and hence should be included in the IR theory of the BLL. Further, $W_\gamma$ will be independent of $x,y$ within the IR theory. 

Thus we have the full set of 3 independent winding numbers expected for an emergent 2-form  $(
\UU(1))_2$ symmetry, in addition to the $\LU(1)$ loop group symmetry.  There is a mixed anomaly between these two symmetries.   To see this, note that in the presence of an $\LU(1)$ gauge field $A_\mu$, the currents of the 2-form symmetry become
\be
{\mathcal J}^{\mu \nu \lambda} = \frac{1}{2\pi} \epsilon^{\mu\nu\lambda\kappa}\left( \partial_\kappa \phi - A_\kappa \right) 
\ee
Clearly then the ${\mathcal J}^{\mu \alpha\beta}$ are no-longer conserved but satisfy 
\be
\partial_\mu {\mathcal J}^{\mu \alpha \beta} = \frac{1}{4\pi}  \epsilon^{\alpha\beta\lambda\kappa} F_{\lambda \kappa} 
\ee
where $F_{\lambda \kappa} = \partial_\lambda A_\kappa - \partial_\kappa A_\lambda$ is the field strength for the $\LU(1)$ gauge field. 

Ref.~\cite{lake2021bose} argued that the BLL has zero phase stiffness associated with  the global $U(1)$ symmetry but a non-zero Drude weight, {\em i.e}, the physical electrical conductivity has a delta-function  peak at zero frequency: 
\be
\sigma(\omega) = D \delta(\omega) + ...
\ee
The presence of this delta function can be understood within the framework described in this paper, as we elaborate below.  First let us define the vector field $u_k$ (with $k = x,y, \gamma$) through
\be
{\mathcal J}^{0ij}  =  \epsilon^{ijk}  u_k.
\ee
Similar to the other examples discussed in the main text, the anomaly equation implies that the  $\LU(1)$ charge density $n(x)$ (with $x = (\bf x, \gamma)$) satisfies the commutation relation 
\be
[n(x), u_k(x')] = -\frac{i}{2\pi} \frac{ \partial}{\partial x_k'} \delta^{3}(x - x') 
\ee
Consider now the generalized Gibbs ensemble 
\be
\rho = e^{-\beta\left({\hat{H}} - \int_\gamma \mu_\gamma \hat{N}(\gamma) - \int d^3 x h^k(x) u_k(x) \right) }
\ee
where $h^k(x)$ is the thermodynamic conjugate to $u_k(x)$ and satisfies $\partial_i h^i  = 0$, and $N_\gamma d\gamma$ is the total $\LU(1)$ charge at point $\gamma$.  Repeating the logic from the main text leads to an $\LU(1)$  current 
\be
\label{eq:bll_current}
j^k(x) \simeq \frac{h^k(x)}{2\pi} 
\ee
Next consider an electric field impulse along the $x$-direction: 
\be
E_x(t) = {\cal E} \delta(t) 
\ee
with ${\cal E}$ independent of $x$. From the anomaly equation this will lead to a change of $u_x$: 
\be
\label{eq:bll_deltau}
\Delta u_x = \frac{{\cal E}}{2\pi}
\ee
We may now relate this to the change of $h_x$ using the thermodynamic susceptibility of the $2$-form conserved charge. To keep things simple we will consider the limit of a patch diagonal action, ignoring Landau parameters that couple together different patches. In that case the (Euclidean) BLL action has the form
\be
\label{eq:S_BLL}
{\cal S}_{BLL} = \int d^2{\bf x} d\tau \frac{d\gamma}{2\pi} \frac{k_B}{4\pi  \eta}  \left(v^{-1} (\partial_\tau \phi_\gamma)^2 + v [\mathbf{w}(\gamma) \cdot \nabla \phi_\gamma]^2 \right) 
\ee
where $\mathbf{w}(\gamma)$ is the unit vector normal to the Bose surface at the point $\gamma$, and $\nabla \phi_\gamma$ denotes the gradient of $\phi_\gamma$ with respect to $x$ and $y$. If we make a change of variables $\nabla \phi_\gamma \to \nabla \phi_\gamma + 2\pi \Delta \mathbf{u}$ in \eqnref{eq:S_BLL}, we see that this is equivalent to shifting the conjugate field $\mathbf{h}$ by
\begin{equation}
\Delta \mathbf{h} = \frac{k_B v}{\eta} [\mathbf{w}(\gamma) \cdot \Delta \mathbf{u}] \mathbf{w}(\gamma).
\end{equation}
In other words, the inverse susceptibility matrix $\chi^{-1}$, defined by $\Delta h^i = (\chi^{-1})^{ij} \Delta u_j$, has components
\begin{equation}
(\chi^{-1})^{ij} = \frac{k_Bv}{\eta} w^i(\gamma) w^j(\gamma).
\end{equation}
Although this is a degenerate matrix (i.e.\ it has a zero eigenvalue), it is not the \emph{zero} matrix, and as a consequence the dissipationless current is not suppressed (that is, there is no critical drag). Indeed, using the expression \eqnref{eq:bll_current} for the current, as well as the shift $\Delta \mathbf{u}$ for an electric field impulse in the $x$ direction given by \eqnref{eq:bll_deltau}, we find that the $\mathrm{LU}(1)$ current is given by
\begin{align}
\label{eq:lu1current}
j^x(\gamma) &= \frac{k_B v \mathcal{E}}{(2\pi)^2} \cos^2 \gamma \nonumber, \\
j^y(\gamma) &= \frac{k_B v \mathcal{E}}{(2\pi)^2} \cos \gamma \sin \gamma.
\end{align}
Therefore, integrating over $\gamma$ to get the total electric current, we find that it is in the $x$ direction and given by
\begin{equation}
j^x_{\mathrm{tot}} = \frac{k_B v \mathcal{E}}{4\pi \eta}.
\end{equation}
This gives a Drude weight $D = k_B v / (4\eta)$ in agreement with the result in Ref.~\cite{lake2021bose}.

\bibliography{ref-autobib,ref-manual}

\end{document}